\documentclass[sn-mathphys,Numbered,pdflatex]{sn-jnl}
\usepackage{graphicx}%
\usepackage{multirow}%
\usepackage{amsmath,amssymb,amsfonts}%
\usepackage{amsthm}%
\usepackage{mathrsfs}%
\usepackage[title]{appendix}%
\usepackage{xcolor}%
\usepackage{textcomp}%
\usepackage{manyfoot}%
\usepackage{booktabs}%
\usepackage{algorithm}%
\usepackage{algorithmicx}%
\usepackage{algpseudocode}%
\usepackage{listings}%
\usepackage{silence}
\usepackage[caption=false]{subfig}

\newtheorem{example}{Example}%
\newtheorem{remark}{Remark}%

\usepackage{amssymb}
\usepackage{amsthm}
\usepackage{newlfont}
\usepackage{stmaryrd}
\usepackage{mathrsfs}
\usepackage{mathtools}
\usepackage{euscript}
\usepackage{graphicx}
\usepackage{enumerate}
\usepackage{todonotes}
\usepackage{color}
\usepackage{float}
\usepackage{subfig}
\usepackage{cancel}
\usepackage{tikz}
\usetikzlibrary{decorations.markings}
\usepackage{calc}
\usepackage{enumitem}
\usepackage{empheq}

\newsavebox\mytikz
\newlength\tikzheight

\def\MarkLt{4pt}
\def\MarkSep{2pt}
\tikzset{
  TwoMarks/.style={
    postaction={decorate,
      decoration={
        markings,
        mark=at position #1 with
          {
              \begin{scope}[xslant=0.2]
              \draw[line width=\MarkSep,white,-] (0pt,-\MarkLt) -- (0pt,\MarkLt) ;
              \draw[-] (-10*0.5*\MarkSep,-5*\MarkLt) -- (-10*0.5*\MarkSep,5*\MarkLt) ;
              \draw[-] (0.5*\MarkSep,-\MarkLt) -- (0.5*\MarkSep,\MarkLt) ;
              \end{scope}
          }
       }
    }
  },
  TwoMarks/.default={0.5},
}
\usetikzlibrary{patterns,decorations.markings,backgrounds}
\usetikzlibrary{decorations.pathmorphing,shapes.arrows}
\usetikzlibrary{positioning}
\usepackage{pgfplots}
\pgfplotsset{compat =1.18, width = 16 cm, height = 9 cm}
\usepgfplotslibrary{fillbetween}

\usepackage{tikz}
\usepackage{graphicx} 
\usepackage{pgf}
\usetikzlibrary{positioning,fit,calc}
\usetikzlibrary{arrows,automata}
\usepackage{wrapfig}
\usepackage{amscd}
\usepackage{hyperref}
\usepackage{cancel}
\usepackage{color}
\usepackage{orcidlink}
\usepackage{hyperref}
\usepackage{hhline}
\usepackage{import}
\usepackage{changes}
\date{\today}

\let\al=a 

\newcommand{\be}{\begin{equation}}
\newcommand{\en}{\end{equation}}

\def\al{a}

\usepackage{soul}

\newcommand{\bbN}{{\mathbb N}}

\newcommand{\bbR}{{\mathbb R}}

\newcommand{\opunit}{\text{1}\kern-0.22em\text{l}}

\newcommand{\id}{\textrm{d}}

\DeclareMathAlphabet{\mathpzc}{OT1}{pzc}{m}{it}

\let\oldsqrt\sqrt
\def\sqrt{\mathpalette\DHLhksqrt}
\def\DHLhksqrt#1#2{%
	\setbox0=\hbox{$#1\oldsqrt{#2\,}$}\dimen0=\ht0
	\advance\dimen0-0.2\ht0
	\setbox2=\hbox{\vrule height\ht0 depth -\dimen0}%
	{\box0\lower0.4pt\box2}}

\let\al=a 

\let\be=\beta

\DeclareMathAlphabet{\mathpzc}{OT1}{pzc}{m}{it}

\def\bea{\begin{eqnarray}}
\def\eea{\end{eqnarray}}
\def\ba{\begin{array}}
\def\ea{\end{array}}

\begin{document}

\title{Phase diagram and specific heat of a nonequilibrium Curie--Weiss model}

\author[1]{\fnm{Aaron} \sur{Beyen} \orcidlink{0000-0002-4341-7661}}


\author[1]{\fnm{Christian} \sur{Maes} \orcidlink{0000-0002-0188-697X}}

\author[2]{\fnm{Irene} \sur{Maes} \orcidlink{0000-0001-5734-956X}}

\affil[1]{\orgdiv{Department of Physics and Astronomy}, \orgname{KU Leuven}}
\affil[2]{\orgdiv{Department of Mathematics}, \orgname{KU Leuven}}



\abstract{
Adding activity or driving
to a thermal system may modify its phase diagram and response functions. We study that effect for a Curie-Weiss model where the thermal bath switches rapidly between two temperatures. 
The critical temperature moves with the nonequilibrium driving, opening up a new region of stability for the paramagnetic phase (zero magnetization) at low temperatures. Furthermore, 
phase coexistence between the paramagnetic and ferromagnetic phases becomes possible at low temperatures. 
\\
Following the excess heat formalism, we calculate the nonequilibrium thermal response and study its behaviour near phase transitions. Where the specific heat at the critical point makes a finite jump in equilibrium (discontinuity), it diverges once we add the second thermal bath.  
Finally, (also) the nonequilibrium specific heat goes to zero
exponentially fast with vanishing temperature, realizing an extended Third Law.}


\maketitle


\section{Introduction}
\begin{figure}[H]\center
			\includegraphics[scale=0.33]{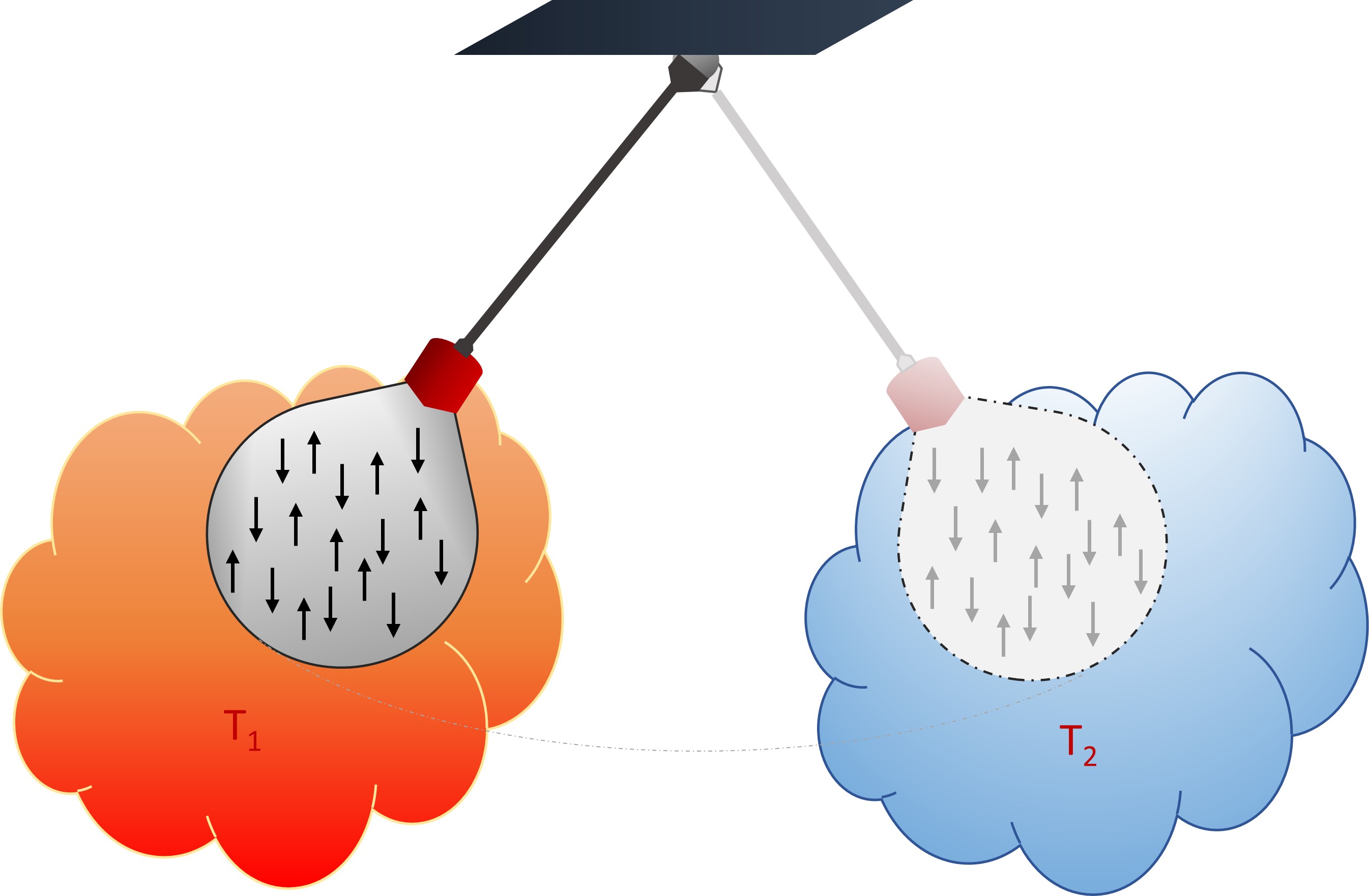}  
		\caption{\small{Spins in alternating contact with two reservoirs at different temperatures. There is no ``average'' temperature effectively describing a detailed balance dynamics of the magnetization.  Courtesy of Faezeh Khodabandehlou.}}\label{2t}
	\end{figure}

Little is understood concerning the changes in the phase diagram for a macroscopic system subject to nonequilibrium driving.  Even less is known about corresponding modifications in critical behaviour and about their thermal markers, \cite{neqphasetransition}. That includes computing the critical exponent for a possible divergence of heat capacities at the (new) critical points, \cite{criticalexponents}.  
The present paper studies those questions for a nonequilibrium mean-field Ising model in the form of a two-temperature Curie-Weiss model, \cite{friedli_velenik_2017}.
Multiple-temperature models have been the subject of nonequilibrium modelling for a long time.  After all, it often makes physical sense to separate the total system into different subsystems (photons and electrons, or, electronic and lattice degrees of freedom) at different temperatures, \cite{Eesley1986GenerationON, masterequationtwotemp, anisotropic}. 
That has been an interesting ingredient for the study of relaxation and transport, and it has also been a subject in the context of steady-state thermodynamics and hydrodynamics, \cite{C_Maes_1991}.  Two-temperature Ising models have been investigated in various guises as well, \cite{superstatistical, criticaluniversality, conservedydynamics, ising1d, energyflux, energyspectrum, Lavrentovich_2012}.\\  
For understanding new many-body physics, here nonequilibrium calorimetry, it is not unusual to start with a mean-field treatment.  Yet, for steady spatially extended multiple-temperature spin systems, it is natural to drive the system from installing contacts with different thermal baths at the spatial \emph{boundaries}.  That is not an option for mean-field approaches and we therefore choose a \emph{temporal} driving.  More precisely, we take the system to alternate between two thermal baths, with inverse temperatures  $\beta_{1}$ and $\beta_{2}$. The spin dynamics then follow the rules of a Markov jump process (kinetic Ising model \cite{Dattagupta}) where the heat bath switches between the two temperatures at random times with rate $r$; see Fig.~\ref{2t}. For mathematical simplicity, we consider the limit $r\uparrow \infty$ so that the spin-flip rate becomes the sum of the two transition rates,  separately satisfying detailed balance with respect to inverse temperature $\beta_1$, respectively $\beta_2$, with corresponding heat fluxes.\\

The main topics of the present paper are on the one hand the exploration of the nonequilibrium phase diagram, and on the other hand, the computation of nonequilibrium thermal responses in the form of heat capacities.  All that is made specific for a two-temperature Curie-Weiss model.\\
Here is a summary of the findings. First and foremost, the phase diagram as labelled by the stationary magnetization changes compared to equilibrium. New regions of stability open up for macroscopic behaviour that is unstable under equilibrium. For example, even at low temperatures, zero magnetization is still stable, indicating the paramagnetic phase in coexistence with a ferromagnetic phase. Furthermore, the critical values for the temperature and for the quartic spin coupling move with the degree of nonequilibrium.
\\Secondly, the nonequilibrium specific heat is calculated in the thermodynamic limit.  We use the excess heat method applied to the macroscopic dynamics. We study its behaviour through the obtained phase diagram. The specific heat diverges at the phase transitions with critical exponent $\alpha = 1$. That starkly contrasts with the equilibrium case, where the specific heat only makes a finite jump \cite{friedli_velenik_2017}. Finally, we check that the computed nonequilibrium specific heat decreases exponentially to zero at low temperatures, realizing a Third Law, \cite{nernst_heat_example}.\\
\\
\textbf{The plan of the paper} is as follows. In Sections \ref{section model} and \ref{section dynamics}, we start with a detailed explanation of the model at finite $N$ and we derive the macroscopic dynamics ($ N \uparrow \infty$), respectively. 
Next, Section \ref{section phase diagram} describes the possible phases and phase transitions of the (nonequilibrium) system for zero magnetic field. We present the nonequilibrium specific heats and we discuss their properties in Section \ref{section calculation heat capacity}.  The detailed derivation of the specific heat formula is given in Section \ref{mpe}. 

\section{Finite model}\label{section model}
To start, the two-temperature Curie--Weiss model is described for a finite set of spins, and we explain how to calculate the nonequilibrium heat capacity. 
\subsection{Setup}
We have a numbered collection of spins $\sigma = (\sigma_{1}, \sigma_{2},...,\sigma_{N})$ for (eventually) large $N$ with $\sigma_{i} = \pm 1$.
In the Curie-Weiss format of the Ising model \cite{friedli_velenik_2017}, the total energy of the spins equals
\begin{equation}\label{energy and psi}
    E(\sigma) := - J N \psi(m^N(\sigma)) \qquad \text{ with } \,\psi(m) := \frac{1}{2} m^{2} + h m + \frac{g}{4} m^{4}
\end{equation}
only depending on the magnetization
\begin{equation*}
    m^{N}(\sigma) :=\frac{1}{N} \sum_{i=1}^{N} \sigma_{i}
\end{equation*}
The parameters $J>0$ (ferromagnetic coupling coefficient), $h$ (magnetic field), and  $g > 0$ \color{black} (quartic interaction) are to be considered as effectively similar  to their counterparts in short-range
Ising-type models on highly connected (high-dimensional) graphs; see \cite{friedli_velenik_2017, Kochma_ski_2013}. 
\\ We are mostly interested in the symmetric case $h = 0$.  The quartic term with coupling $g$ has been used for modelling quartic interactions, \cite{quartic, quartic2, quartic3}  and also appears in the Landau theory of phase transitions \cite{Landau:1980mil,LGtheory}. We include it here mainly for completeness and to allow for more complex phase diagrams (even in equilibrium). For example, when $g \neq 0$, a coexistence appears between ferro- and paramagnetic phases. Going nonequilibrium also modifies the critical exponent when the specific heat diverges for some $g$.
\\
\\
The dynamics follows a kinetic Ising model \cite{Dattagupta, Liggett1985InteractingParticleSystems}, where for the total rate, 
we take a sum of two spin flip rates, each corresponding to a thermal bath. The spin-flip rates separately satisfy local detailed balance (see \cite{ldb}) with respect to its thermal reservoir, one at inverse temperature $\beta_1\geq 0$ and the other at $\beta_2\geq 0$. For definiteness, we take $\beta_2 \geq \beta_1$, {\it i.e.}, the second bath is at a lower temperature.
\\ The only allowed transitions are between $\sigma = (\sigma_{1}, \sigma_{2},..., \sigma_{k},...,\sigma_{N})$ and $\sigma^{k} = (\sigma_{1}, \sigma_{2},..., -\sigma_{k},...,\sigma_{N})$, for some $k \in \{1,2,...,N\}$, {\it i.e.}, one spin flip at a time. 
The (finite) two-temperature Curie-Weiss dynamics is then the Markov process on $K_N:= \{-1,+1\}^N$ with  rates $c(\sigma,k)$ for flipping from $\sigma \to \sigma^k$:
\begin{eqnarray}\label{sumr}
 c(\sigma, k) &=& c_1(\sigma, k) + c_2(\sigma, k),\\
 c_a(\sigma,k) &=&   \frac{\nu}{(J\beta_{a})^{n}}\, e^{\frac{\beta_{a}}{2}(E(\sigma)-E(\sigma^k))},\qquad a=1,2 \label{alra} 
\end{eqnarray}
The reference frequency $\nu > 0$ is added to keep the correct units in \eqref{alra}.  It sets the time scale but we are interested only in the asymptotic behaviour $t \uparrow \infty$ (after taking the thermodynamic limit), which can not be influenced by $\nu$. 
More importantly, a temperature-dependent activation is added in the prefactor of \eqref{alra}, where we recognize an Arrhenius (for $n=0$) or Eyring (for $n=1$) power-law, \cite{arrhenius_svante_1889_1448930, Eyring}. In chemical kinetics, these prefactors are usually interpreted as the collision frequency of the molecules in question, increasing with temperature. For spins, there is a similar idea where higher energy/temperature induces more activity ($n\geq 0$).  In thermal equilibrium with $\beta_1=\beta_2$, such prefactors are not important but they can influence the dynamics when out of equilibrium.  Kinetic differences (here, for example, already for different $n$) do matter for nonequilibrium phase diagrams and responses.  In this paper, to be specific, we compare the case with ($n = 1)$ and without $(n = 0)$ temperature--dependence in the prefactor. Obviously, more general transition rates exist for \eqref{alra}, but we stick to the current choice. \\

The dynamics \eqref{sumr}--\eqref{alra} can be seen as the limit $r\uparrow \infty$ of a Curie-Weiss model where the temperature switches randomly at rate $r$ between the two values; see Fig.~\ref{2t}. Indeed, the case of finite $r$ can be modeled by introducing a dichotomous noise $\eta_t \in \{0,1 \}$, which switches between $0$ and $1$ at rate $r$ \cite{dichnoise}.
Then,
\begin{equation}\label{finite r}
    \tilde{c}(\sigma,k) = \eta_t \ c_1(\sigma,k) + (1-\eta_t) \ c_2(\sigma,k)
\end{equation}
and the Markov process $(\eta_t, \sigma_t)$ gets correlated. However, in the limit $r \uparrow \infty$, they decorrelate, $\langle \eta_t \rangle = \frac{1}{2}$, and
\begin{equation*}
   \langle \tilde{c}(\sigma,k) \rangle \to \frac{c_1(\sigma,k) + c_2(\sigma,k)}{2}  
\end{equation*}
where the factor $\frac{1}{2}$ can be absorbed in the time parameter.  That averaging obviously does not apply for the temperatures, due to the nonlinearities in \eqref{sumr}--\eqref{alra} and the absence of global detailed balance.
    
\subsection{The Markov spin-flip dynamics}\label{markov spin flip dynamics}
The backward generator of the Markov process under consideration is $L^{N} = L_{1, N} + L_{2, N}$ where, for $\sigma,\sigma'\in K_N$,
\begin{equation}\label{La}
     L_{a,N}(\sigma, \sigma') = \left\{
    \begin{array}{ll}
        c_{a}(\sigma,k) & \mbox{if } \sigma' = \sigma^k \text{ for some } k \in \{1,2,...,N\} \\
        - \sum_k c_{a}(\sigma, \sigma^k) & \mbox{if } \sigma = \sigma' \\
        0 & \mbox{ else}
    \end{array} \right.
\end{equation}
is the backward generator $L_{a, N}$ of the Markov jump process when only coupled to temperature bath $a=1,2$. The Master equation is
\begin{equation}\label{master equation l dagger}
    \frac{\id \rho_t}{\id t} = \rho_t L^N = L^{N \dagger} \rho_t
\end{equation}
where $L^{N \dagger}$ (forward generator) is the transpose of $L^N$, and the probability  $\rho_t$ is a row vector of the $N$ spins. As a reminder, to settle the notation,
\begin{gather}
    L_{a,N} f(\sigma) = \sum_{i = 1}^N c_a(\sigma,i) [f(\sigma^i)-f(\sigma)]  \label{ll} \\
     \frac{\id}{\id t} \langle f(\sigma_t) \rangle = \langle L^N f(\sigma_t) \rangle \qquad \langle f(\sigma_t) | \sigma_0 = \sigma \rangle = e^{tL^N } f\,(\sigma) \label{properties L 2}
\end{gather}
for all functions $f:K_N\rightarrow \bbR$. 
\\Finally, there is a unique stationary distribution $\rho_N^s$, stationary solution of \eqref{master equation l dagger}, satisfying
\begin{equation*}
     \rho_N^s L^N = 0
\end{equation*}
  That allows to introduce the stationary expectation $\langle \cdot \rangle_N^s$ and
\begin{eqnarray}
    \langle f\rangle^s_N & = &\sum_{\sigma\in K_N} f(\sigma)\, \rho_N^s(\sigma), \qquad \langle L^{N}g\rangle_N^s = 0\label{ts}
\end{eqnarray}
for arbitrary functions $f,g$ on $K_N$.  Obviously, $\rho_N^s $ depends on the two temperatures.\\

The thermal interpretation (below) is inspired by the case of finite $r$; see Fig.~\ref{2t}. 
The heat sent  to the $a$-th thermal bath during the flipping of the spin at $k$ equals
\begin{equation}\label{he1}
q_{a,N}(\sigma,k) = \frac 1{\beta_a}\log \frac{c_a(\sigma,k)}{c_a(\sigma^k,k)} = E(\sigma) - E(\sigma^k), \quad a=1,2
\end{equation}
When the configuration is $\sigma$, the instantaneously expected heat flux (or power) $P_{a,N}(\sigma)$ to the heat bath $a$ is, therefore, 
\begin{equation}\label{hfn}
P_{a,N}(\sigma) 
= \sum_k c_a(\sigma,k)\, q_{a,N}(\sigma,k) = -L_{a,N}E(\sigma)
\end{equation}
In the last line, we used the definition \eqref{La} of the backward generator $L_{a, N}$.
Finally, from \eqref{hfn}, the stationary heat flux $j_{a,N}^s$ to the $a$-th reservoir, per number of spins, becomes
\begin{equation}\label{fheat current formula}
    j_{a,N}^s = \frac{1}{N} \langle P_{a,N} \rangle_N^s =  -\frac 1{N}\langle L_{a,N}E\rangle_N^s = \frac 1{N}\langle L_{b,N}E\rangle_N^s = -j_{b,N}^s, \quad a\neq b
\end{equation}
where we used $\langle L_{a,N}E\rangle_{N}^{s} = -\langle L_{b,N}E\rangle_{N}^{s}$ when $a\neq b$ (see \eqref{ts}) and where the 
energy function $E$ needs to be taken from \eqref{energy and psi}.\\

\subsection{Excess heat, quasipotential and heat capacity}\label{fhe}

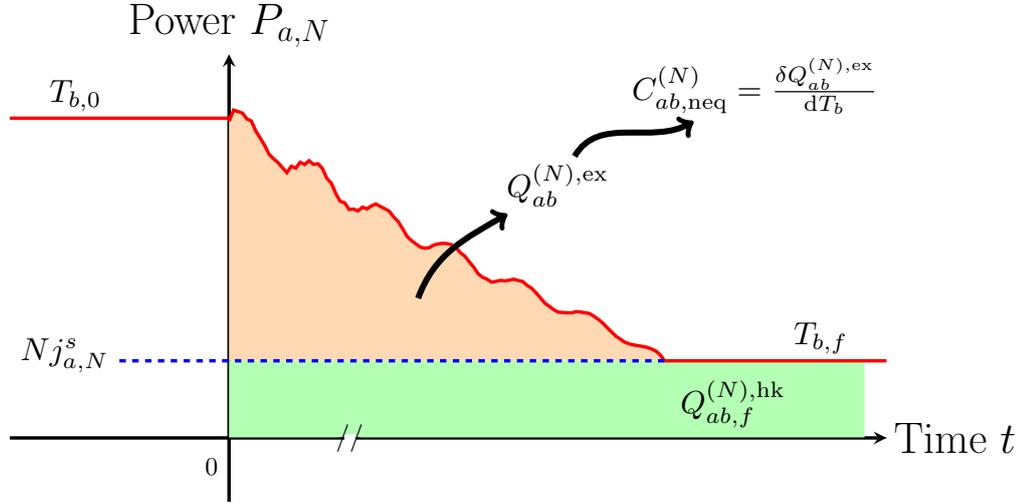
\begin{figure}[H]
    \centering
    \scalebox{0.8}{
    \begin{tikzpicture}[]
    \begin{axis}[axis lines = middle, xlabel = {Time $t$}, ylabel = {Power $P_{a,N}$}, xmin = -100, xmax = 300, ymin = -10, ymax = 60, x axis line style={ultra thick}, y axis line style={ultra thick},
        xtick ={\empty},
        ytick = {\empty},samples=100,x label style={anchor=west, font = \huge}, y label style={anchor=south, font = \huge}]
        \addplot[color = red, domain =-100:0, ultra thick]{50};
        \addplot[name path = A,color = red,ultra thick, domain =0:201]{(sqrt((50-10)^2-x^2/25)+2.5*sin(deg(x/5)))*e^(-0.03*x/5)+10+0.9*rand*e^(-0.08*x/5)};
        \addplot[name path = B,color = blue, dashed, ultra thick,domain =-50:200]{12.06};
         \addplot[name path = C,color = black, ultra thin,domain =0:300]{0};
         \addplot[name path = D,color = red, domain =199:300, ultra thick]{12.06};
        \addplot[orange!30] fill between[of= A and B,soft clip={domain=0:200}];
        \addplot[green!30] fill between[of= B and C,soft clip={domain=0:290}];
        \addplot[green!30] fill between[of= D and C,soft clip={domain=0:290}];
        \node [above] (2) at (150,35) {\scalebox{1.5}{$Q^{(N),\text{ex}}_{ab}$}};
        \node at (240,55) (3) {\scalebox{1.5}{$C^{(N)}_{ab, \text{neq}} = \frac{\delta  Q^{(N),\text{ex}}_{ab}}{\id  T_b}$}};
        \node [above] at (-70, 50) {
        \scalebox{1.5}{$T_{b, 0}$}};
        \node [above] at (270, 12) {\scalebox{1.5}{$T_{b,f}$}};
       \node (1) [above] at (85,19.5) {};
       \draw[->, line width = 1 mm]  (1) [out=70, in=-150] to  (2);
       \draw[->, line width = 1 mm]  (2) [out=60, in=-150] to  (3);
       \node (4) [above] at (230,0.5) {\scalebox{1.5}{$Q_{ab, f}^{(N),\text{hk}}$}};
       \draw (50,0) -- node[fill=white, ultra thick, rotate=0,inner xsep=-4 pt, inner ysep = -5]{\scalebox{1.3}{//}} (60,0);
       \node [above] at (-8, -7) {
        \scalebox{1.2}{0}};
        \node (4) [above] at (-75,9) {\scalebox{1.5}{$N j_{a,N}^s$}};
    \end{axis}

\end{tikzpicture}
}

    \caption{Cartoon of the heat flux or power $P_{a,N}$ as a function of time for a relaxation process when changing the temperature $T_{b,0}\rightarrow T_{b,f} = T_{b,0} + \id T_{b}$.  There are two sources of heat (time-integrated power): the excess heat $Q^{(N),\text{ex}}_{ab}$ and the housekeeping heat $Q^{(N), \text{hk}}_{ab}$. In the quasistatic limit, the nonequilibrium heat capacity is given by $C^{(N)}_{ab, \text{neq}} = \frac{\delta  Q^{(N),\text{ex}}_{ab}}{\id  T_b}$; see \eqref{fheat capacity formula} for more precision.   (Made in Tikz).}
    \label{excess heat tikz}
\end{figure}

The definition of nonequilibrium heat capacity follows the excess heat framework discussed in \cite{nernst_heat_example,nonequilibrium_calorimetry,nonequilibrium_steady_states}. To be more self-contained, we use Fig.~\ref{excess heat tikz} to guide the reader.

One imagines that the system is in a steady nonequilibrium condition before time $t=0$.  The temperature of the two heat baths is fixed, $T_{b,0}$ being the temperature of bath $b$.  Obviously, the spin configuration $\sigma_0=\sigma$ at time $t=0$ is random, and follows distribution $\rho^s_N$ at $T_{b,0}$.  At time $t=0$ we slightly change the temperature $T_{b,0} \longrightarrow T_{b,f} =  T_{b,0} + \id T_{b}$.  There will now start a relaxation to a new steady condition; the  $\rho^s_N$ for $T_{b,0}$ will converge to the $\rho^s_N$ for $T_{b,f}$.  That takes time of course, while the system is exposed to $T_{b,f}$ in heat bath $b$ while the temperature in heat bath $a$ has not changed.  At times $t > 0$, the power dissipated in heat bath $a$ is $P_{a,N}(\sigma_t)$ with Markov process conditional expectation $\langle P_{a,N}(\sigma_t)\,|\,\sigma_0=\sigma\rangle_N$.  Asymptotically, for $t\uparrow\infty$ the expected dissipated power converges to $Nj_{a, N}^s$ for temperature $T_{b,f}$ in heat bath $b$  (see Fig. \ref{excess heat tikz}).  Therefore, the excess heat to heat bath $a$ is estimated by the time integral of the difference,
 \begin{equation}\label{fquasipotential formula}
    V^{(N)}_{a}(\sigma) := \int_{0}^{\infty} \id t\,[\langle P_{a,N}(\sigma_t)|\sigma_0 = \sigma \rangle_N - Nj_{a,N}^{s}] 
\end{equation}
Note that the integral is converging since 
\[
\lim_{t\uparrow \infty} \langle P_{a,N}(\sigma_t)\,|\,\sigma_0=\sigma\rangle_N  = Nj_{a,N}^s
\]
exponentially fast, uniformly in $\sigma$, as a standard consequence of the Perron–Frobenius theorem \cite{perron-frob}. Similarly,  
\begin{equation}\label{condition V discrete}
    \langle V^{(N)}_{a} \rangle^s_N=0
\end{equation}
and the dynamics in \eqref{fquasipotential formula} and the stationary distribution are all at temperature $T_{b,f}$.\\
The time-integral \eqref{fquasipotential formula} makes $V^{(N)}_{a}(\sigma)$ a measure of the excess heat, indicated by $Q^{(N),\text{ex}}_{ab}$ in Fig.~\ref{excess heat tikz}.  It is the heat that is in excess with respect to the {\it housekeeping} heat $Q^{(N),\text{hk}}_{ab}$ also indicated in Fig.~\ref{excess heat tikz}, and is entirely due to the relaxation from the original steady nonequilibrium condition to the final one. It is indeed important to realize that dissipation is already present even before any parameter, like the temperature, is changed.  The nonequilibrium heat capacity $C^{(N)}_{ab}$ can therefore be defined as the thermal response of $Q^{(N),\text{ex}}_{ab}$ to a change in temperature in heat bath $b$.\\
A more precise analysis takes the previous heuristics through a quasistatic limit (which is the reason we did not immediately identify $Q^{(N),\text{ex}}_{ab}$ with $V^{(N)}_{a}$) but the idea is unchanged:\\
For each finite $N$, the nonequilibrium heat capacity $C_{ab}^{(N)}$ from measuring the excess heat flux from bath $a$ due to a change in the temperature of reservoir $b$ ($a,b=1,2$) equals 
\begin{equation}\label{fheat capacity formula}
    C_{ab}^{(N)} = -\left\langle \frac{\partial V_a^{(N)}}{\partial T_{b}} \right\rangle_{N}^s = k_B \beta_{b}^2 \left\langle \frac{\partial V_a^{(N)}}{\partial \beta_{b}} \right\rangle_{N}^s
\end{equation}
The sum
\begin{equation*}
    C_b^{(N)} = C_{ab}^{(N)} + C_{bb}^{(N)}, \qquad b \neq a
\end{equation*}
measures the thermal susceptibility of the total environment (both thermal reservoirs combined) to changes in reservoir temperature $T_b$.
Experimentally, a nonequilibrium heat capacity can be obtained from AC-calorimetry measurements, see \cite{Dolai_2023}. 
For computational purposes, it is interesting to observe that the integral equation \eqref{fquasipotential formula} can be reduced to the following equivalent Poisson equation for $V^{(N)}_{a}$,
\begin{align}
L^{N} V^{(N)}_{a} &=  N j_{a,N}^s - P_{a,N} \label{Poisson eq finite N}\\
& =  L_{a,N}E - \langle L_{a,N}E\rangle^s_{N} \nonumber \\
& =  L_{a,N}E + \langle L_{b,N}E\rangle^s_{N},\qquad b\neq a \label{pep}
\end{align}
where we used that $\langle L_{a,N}E\rangle_{N}^{s} = -\langle L_{b,N}E\rangle_{N}^{s}$ when $a\neq b$ in the last line; see \eqref{ts}.\\
We see immediately from \eqref{pep} that 
\begin{equation*}
L^{N}(V_1^{(N)}+V_2^{(N)})= L^{N}E
\end{equation*}
which implies that the sum always satisfies
\begin{equation*}
V_1^{(N)} + V_2^{(N)} = E - \langle E\rangle^{s}_{N}
\end{equation*}
which motivates also to call $V_a^{(N)}$ a {\it quasipotential}.\\

As a simple illustration of the above thermal setup, in Appendix \ref{two} we 
give the explicit solution for the case of only $N=2$ spins $\sigma_1, \sigma_2$. Clearly, that system is not macroscopic but it already gives an interesting nonequilibrium generalization (of many independent and identical copies) of the well-known two-level system, showing e.g. negative heat capacities. It also allows the reader to get acquainted with the excess heat formalism briefly introduced above.

\section{Macroscopic dynamics}\label{section dynamics}
When the number $N$ of spins gets very large, simplifications and greater interest arise, which we explain in this section. In particular, we show here that in the thermodynamic limit, the magnetization follows a first-order equation in time.

\subsection{Propagation of molecular chaos}

Consider the  Bernoulli product probability distributions $\rho_m^{\otimes}$ on $\{+1,-1\}^{\bbN_{>0}}$ with parameter $m\in [-1,1]$, formally defined as
\[
 \rho_m^{\otimes}(\sigma) = \prod_{\ell} \frac{1 + m \sigma_\ell}{2}
 \]
with corresponding expectation $\langle \cdot\rangle_m$; $\langle \sigma_1\sigma_2\ldots\sigma_k\rangle_m = m^k$. Note that for that probability distribution, $\langle m^N(\sigma) \rangle_m = m$, such that by the strong law of large numbers, $m^{N}(\sigma) \rightarrow m$ with probability one.
\\
 In other words, for any continuous function $f: [-1,1]\times \{+1,-1\}^k$ with $k\in \bbN_{> 0}$, 
\begin{equation}\label{wlln}
\lim_N \langle f(m^N(\sigma),\sigma_1,\dots,\sigma_k)\rangle_m = \langle  f(m,\sigma_1,\dots,\sigma_k)\rangle_m
\end{equation}
where we still average over the spins $\sigma_1,...,\sigma_k$ in the last expression. \\

Imagine now that at time zero $t = 0$, the spins are sampled from $\rho_m^{\otimes}$ 
with given magnetization $m\in [-1,1]$.
We claim, as is typical for mean-field systems in the limit $N \uparrow \infty$ \color{black}, that the spins at later times $t\geq 0$  follow a product distribution  $\rho_{m_t}^{\otimes}$ as well 
provided that
\begin{equation}\label{dmdt first time}    
\frac{\id m_t}{\id t} = H(m_t) := H_{1}(m_t)+  H_{2}(m_t),\quad m_0=m
\end{equation}
for 
\begin{equation}\label{Ha}
H_a(m) := \frac{2 \nu}{(J\beta_{a})^n} \Big(\sinh{J \beta_{a} \psi'(m)} - m\, \cosh{J\beta_{a} \psi'(m)} \Big)
\end{equation}
We call that claim the {\it propagation of chaos}, \cite{chaos, molecularchaos}, indicating that
the spins remain statistically independent in the limit $N \uparrow \infty$,\\
To show that propagation of chaos, we will prove that the weak limit of the forward generator $L_{a, N}^{\dagger}$ (transpose of $L_{a, N}$) satisfies
\begin{equation}\label{L dagger rho^k}
    L^+\rho_m^{\otimes} := \lim_N \sum_{a = 1}^2 L^+_{a,N}  \rho_m^{\otimes} =H(m)\,\frac{\partial \rho_m^{\otimes}}{\partial m}
\end{equation}
As a consequence of \eqref{L dagger rho^k}, if indeed $(m_t, t\geq 0)$ is a solution of \eqref{dmdt first time}, then
\[
\frac{\id}{\id t}\rho_{m_t}^\otimes = \frac{\partial \rho_{m_t}^{\otimes}}{\partial m_t} \frac{\id m_t}{\id t} = \frac{\partial \rho_{m_t}^{\otimes}}{\partial m_t} H(m_t)  =   L^\dagger \rho_{m_t}^\otimes
\]
and hence $\rho_{m_t}^\otimes$ is the solution of the (infinite volume) Master equation \eqref{master equation l dagger} (which is the property of propagation of chaos).  Since the solution to the Master equation is unique, it follows that, in the limit $N \uparrow \infty$, we are only interested in a dynamics satisfying \eqref{dmdt first time}.\\

\begin{proof}[Proof of \eqref{L dagger rho^k}]
The forward generator $L_{a,N}^{\dagger}$  acts as 
\begin{eqnarray}\label{L dagger}
\sum_{\sigma} g(\sigma)\,  L_{a,N}^{\dagger}  \rho_m^{\otimes}(\sigma) &=&  \sum_{\sigma} g(\sigma) \sum_{i = 1}^k \left[c_{a}(\sigma^i,i) \rho_m^{\otimes}(\sigma^i) 
- c_a(\sigma,i)  \rho_m^{\otimes}(\sigma) \right] 
\end{eqnarray}
for arbitrary $g(\sigma)=g(\sigma_1,\ldots,\sigma_k), k\leq N$. 
We can use here the asymptotics
\begin{equation}
 c_a(\sigma,k) =   \frac{\nu}{(J\beta_{a})^{n}}\, e^{-J \beta_{a}\sigma_{k} \,\psi'(m^{N}(\sigma))} \left\{ 1 + \frac{\beta_a J}{N} \psi''(m^{N}(\sigma)) + O\left(\frac{1}{N^2} \right)  \right\} \label{ca large N}
\end{equation}
Noting further that
\begin{equation*}
    \rho_m^{\otimes}(\sigma^i) = \rho_m^{\otimes}(\sigma) \frac{1 - m \sigma_i}{1 + m \sigma_i}
\end{equation*}
the right-hand side of \eqref{L dagger} equals
\begin{align*}
 &\frac{\nu}{(J\beta_{a})^{n}}\left \langle g(\sigma)\sum_{i = 1}^k\left[ \frac{1 - m \sigma_i}{1 + m \sigma_i} e^{J \beta_{a}\sigma_{i} \,\psi'(m^{N}(\sigma))} -  e^{-J \beta_{a}\sigma_{i} \,\psi'(m^{N}(\sigma))}\right]  \right \rangle_m + O\left( \frac{1}{N} \right) 
\end{align*}
and hence, up to order $O\left( \frac{1}{N} \right) $,
\begin{align}
&\sum_{\sigma}  \sum_{a = 1}^2 g(\sigma)\,  L_{a,N}^{\dagger}  \rho_m^{\otimes }(\sigma) =\nonumber \\
& \sum_{a = 1}^2\frac{\nu}{(J\beta_{a})^{n}}\left\langle g(\sigma)\sum_{i = 1}^k\left[ \frac{1 - m \sigma_i}{1 + m \sigma_i} e^{J \beta_{a}\sigma_{i} \,\psi'(m)} -  e^{-J \beta_{a}\sigma_{i} \,\psi'(m)}\right]\right\rangle_m  \nonumber \\
&\sum_{a = 1}^2 \frac{\nu}{(J\beta_{a})^{n}}\left\langle g(\sigma)\sum_{i = 1}^k \frac{2\sigma_i}{1+ m\sigma_i}\left[
\sinh{\beta_a J \psi'(m)} - m \cosh{\beta_a J \psi'(m)}\right]
\right\rangle_m = \nonumber \\
& =\sum_{\sigma} g(\sigma)   \sum_{a = 1}^2 H_a(m)\frac{\partial \rho_m^{\otimes}}{\partial m}(\sigma) = \sum_{\sigma} g(\sigma)   H(m) \frac{\partial \rho_m^{\otimes}}{\partial m}(\sigma) \label{proposition rhok}
\end{align}
where we used in the last line the definition \eqref{Ha} of $H_a$  together with the fact that, formally, 
\begin{align*}
    \frac{\partial \rho_m^{\otimes}}{\partial m}(\sigma) = \frac{\partial}{\partial m} \Big(\prod_{i} \frac{1 + m \sigma_i}{2} \Big) = \rho_m^{\otimes}(\sigma) \,\sum_{i} \frac{\sigma_i}{1 + m \sigma_i} 
\end{align*}
Since \eqref{proposition rhok} holds for arbitrary functions $g$ (depending on a finite number of spins), the weak limit \eqref{L dagger rho^k} is proven.
\end{proof}

\subsection{Magnetization dynamics}
The function $H = H_1 + H_2$, defined from \eqref{Ha},  is smooth in $m \in [-1,+1]$ and satisfies 
\begin{equation*}
   H(\pm 1) = H_1(\pm 1) + H_2(\pm 1) = \mp 2 \nu \left( \frac{e^{- J \beta_1 \psi'(\pm 1)}}{(J \beta_1)^n} + \frac{e^{- J \beta_2 \psi'(\pm 1)}}{(J \beta_2)^n} \right)
\end{equation*}
such that $\text{sgn}(H(\pm 1)) = \mp 1$. As a consequence, by looking at \eqref{dmdt first time}, if $m_0 \in [-1,1]$, then also $m_t \in [-1,1]$ for all $t$, and thus ({\it e.g.,} from the Picard–Lindelöf theorem \cite{agarwal1993uniqueness}), the solution $m_t$ to the initial value problem exists and is unique for all times $t\geq 0$.\\ 
In more detail,
\begin{align}\label{dmdt}
     &\frac{\id m_t}{\id t} = H_1(m_t) + H_2(m_t) = \frac{2 \nu \cdot 4^n}{J^n(4 \beta^2-\delta^2)^n} \Biggl[ \nonumber \\ 
     & \left( (\beta - \frac{\delta}{2})^n+(\beta + \frac{\delta}{2})^n \right) \cosh{(\frac{\delta}{2}\psi'(m_t))} \Big(\sinh{\beta  \psi'(m_t)}- m_t \cosh{\beta  \psi'(m_t)} \Big) \nonumber \\
     & + \left( (\beta + \frac{\delta}{2})^n-(\beta - \frac{\delta}{2})^n \right) \sinh{(\frac{\delta}{2}\psi'(m_t))} \Big(m_t \sinh{\beta  \psi'(m_t)}-  \cosh{\beta  \psi'(m_t)} \Big) \Biggr]
\end{align}
Here and in what follows, we use the dimensionless quantities 
\begin{equation}\label{beta, delta}
    \beta =\frac J{2}(\,\beta_1 + \beta_2\,) \qquad \delta = J(\beta_2 - \beta_1) \geq 0
\end{equation}
where the parameter $\delta$ characterizes the degree of nonequilibrium. Even though $\beta,\delta$ are dimensionless, we often refer to them as the mean inverse temperature and inverse temperature difference, respectively. Note that $\beta_1 \geq 0$ implies the constraint $\beta \geq \frac{\delta}{2}$ but $\delta$ need not be small.\\ 

Stationary values $m_*$ of the magnetization satisfy $H_{1}(m_*)+  H_{2}(m_*) =0$ in \eqref{dmdt}, which means for example that $m_*=0$ is always stationary when $h=0$ in \eqref{energy and psi}.  More generally, the stationary magnetization solves
\begin{equation}\label{general ms}
   \frac{\left( (\beta + \frac{\delta}{2})^n-(\beta - \frac{\delta}{2})^n \right)}{\left( (\beta + \frac{\delta}{2})^n+(\beta - \frac{\delta}{2})^n \right)}  \tanh{\frac{\delta}{2} \psi'(m_*)} = -\frac{\sinh{\beta  \psi'(m_*)}-m_* \cosh{\beta  \psi'(m_*)}}{m_* \sinh{\beta  \psi'(m_*)}- \cosh{\beta  \psi'(m_*)}}
\end{equation}
which for $n=0$ in \eqref{sumr} becomes 
\begin{equation}\label{general ms1}
 m_* = \tanh \beta \psi'(m_*)
\end{equation}
independent of $\delta$, and reduces to
\begin{equation}\label{general ms2}
   \delta\, \tanh{\frac{\delta}{2} \psi'(m_*)} = 2\beta\,  \frac{m_* - \tanh{\beta  \psi'(m_*)}}{m_* \tanh{\beta  \psi'(m_*)}- 1}
\end{equation}
for $n=1$. Note further that \eqref{general ms} is symmetric under flipping the sign of $\delta$,
\begin{equation}\label{symmetry ms}
    m_*(\beta,\delta) = m_*(\beta, - \delta)
\end{equation}
which implies that the deviation from the equilibrium magnetization $m_*(\beta,0)$ is of order $\delta^2$ when close to equilibrium.

\section{Phase diagram}\label{section phase diagram}
In this section, the phase diagram for the macroscopic and stationary nonequilibrium Curie-Weiss model \eqref{general ms} is derived for $h=0$.  For comparison,  see {\it e.g.} \cite{friedli_velenik_2017, Kochma_ski_2013} for a systematic review of the equilibrium version.\\
We start with two subsections that treat the cases of values $n=0$ and $n=1$, respectively, as appear in \eqref{Ha} and (originally) in \eqref{alra}.

\subsection{\texorpdfstring{$n = 0$}{n = 0}}\label{subsection n = 0}
\begin{figure}[H]
    \centering
    \subfloat[\centering $g < 1/3$: 1 or 3 solutions for $m_*$.]{{\includegraphics[width=6.cm]{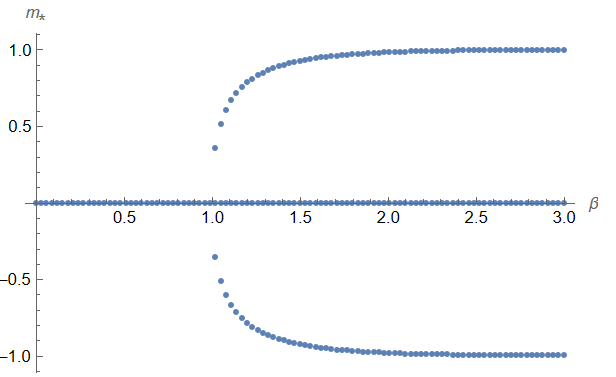}}}%
    \qquad
    \subfloat[\centering $g > 1/3$: 1,3 or 5 solutions for $m_*$.]{{\includegraphics[width=6.cm]{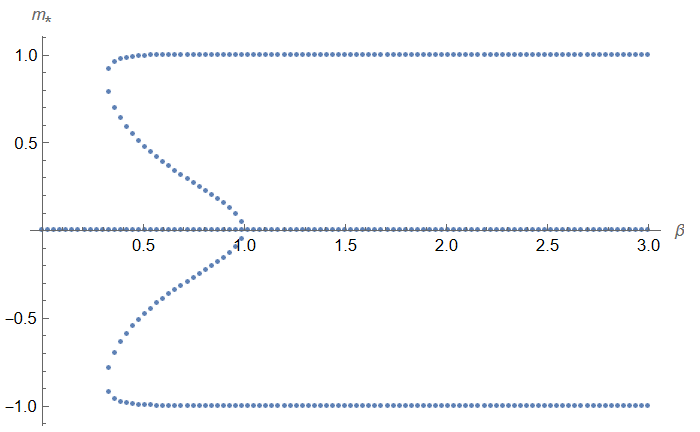}}}%
    \vspace{0.5 cm}
    \caption{Stationary magnetizations $m_*(\beta)$ for different values of $g$. (Made using Mathematica version 13.1.0.0 \cite{Mathematica}.)}%
    \label{ms beta for different g}%
\end{figure}

The stationary values $m_*$ satisfy \eqref{general ms1} (for $h = 0$),
\begin{equation}\label{stationary n = 0}
 \tanh{\left(\beta (m_*+g m_*^{3}) \right)} = m_*,\qquad \beta \geq \frac{\delta}{2}
\end{equation}
where the equality is identical to the equilibrium mean-field equation for the stationary magnetisations of the Hamiltonian \eqref{energy and psi}, but now under the {\it average} inverse temperature $\beta$. However, there is the (nonequilibrium) constraint $\beta \geq \frac{\delta}{2}$ that determines the physically allowed phases.  The dependence on the nonequilibrium parameter $\delta$ is therefore implicitly present.\\
There are  1,3 or 5 solutions to \eqref{stationary n = 0}, depending on the parameters $\beta,g$. Since $h = 0$, the magnetization $m_* = 0$ is always a solution and indicates a paramagnetic phase. The other nonzero solutions $m_* \neq 0$ represent ferromagnetic states. Numerically solving \eqref{stationary n = 0} gives the pairs $(\beta,m_*)$, shown in Fig.~\ref{ms beta for different g}.

\begin{figure}[H]
    \hspace{-0.47 cm}\includegraphics[scale = 0.5]{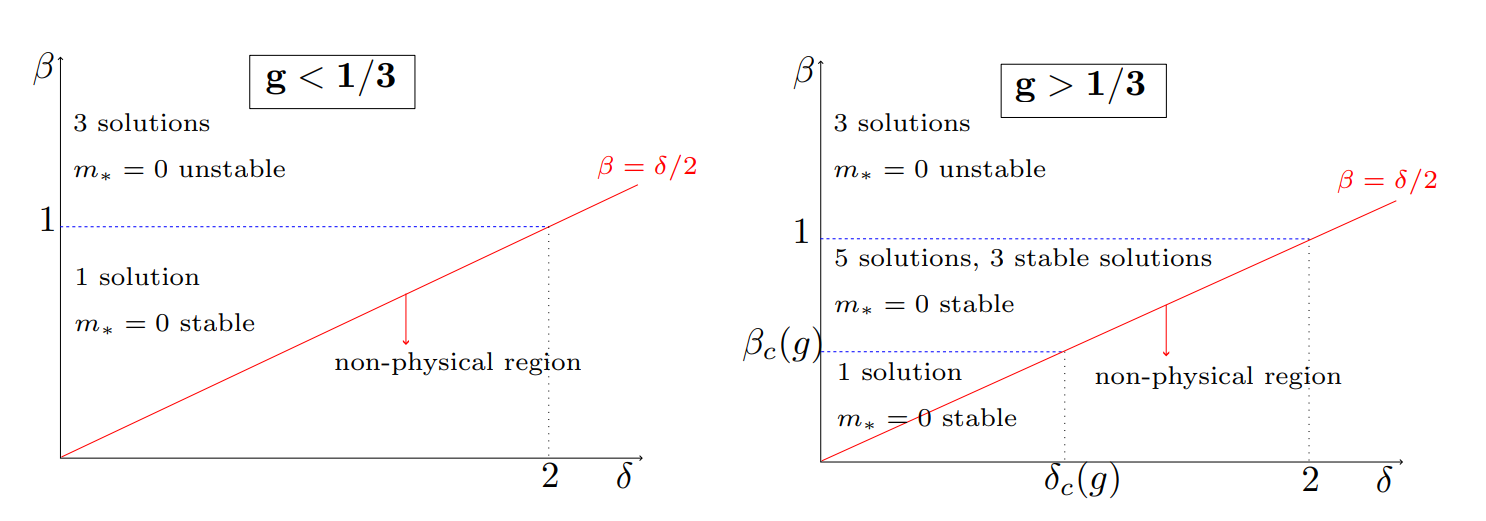}
    \caption{$(\delta, \beta)$ diagram for fixed $g$.}
    \label{phase diagram beta vs delta fixed g n = 0}
\end{figure}

Since 
\begin{equation*}
    H_1'(0)+H_2'(0) = 4(\beta-1)
\end{equation*}
we have  that $m_* = 0$ is stable for $\beta \leq 1$ and unstable for $\beta > 1$ (independent of $g$) with critical inverse temperature $\beta_c = 1$. An expansion of \eqref{stationary n = 0} near $m_* \approx 0, \beta \approx 1$ yields
\begin{equation}\label{near ms = 0 n = 0}
    m_* 
    = \pm \sqrt{\frac{\beta -1}{1/3-g}}  + O \left( (\beta - 1)^{\frac{3}{2}} \right)
\end{equation}
As in equilibrium, $m_* \propto (\beta-\beta_c)^{z}$ with critical exponent $z = \frac{1}{2}$  and $m_* \propto (g_c-g)^{-y}$ where $y = \frac{1}{2}$. Furthermore, for $g < \frac{1}{3}$, the stationary magnetization $m_* \approx 0$ follows $\sqrt{\beta -1}$ for $\beta > 1$, while for $g > \frac{1}{3}$, it grows like $\sqrt{1 - \beta}$ for $\beta < 1$ (implying a second phase transition at  $g_c = \frac{1}{3}$). This behaviour is also shown in Fig. \ref{ms beta for different g}.\\  
Lastly, note that when $\delta \geq \delta_c =2 \beta_c = 2$, then the constraint is $\beta = \frac{\delta}{2} \geq 1 = \beta_c$, such that $m_*=0$ is always unstable for $\delta > 2$. This introduces a critical $\delta_c = 2$ in the phase diagrams.\\
The phase diagrams in Figs.~\ref{g,beta diagram for fixed delta n = 0}--\ref{phase diagram beta vs delta fixed g n = 0} combine all information for $n = 0$.
\begin{figure}[H]
    \hspace{-0.8 cm}\includegraphics[scale = 0.5]{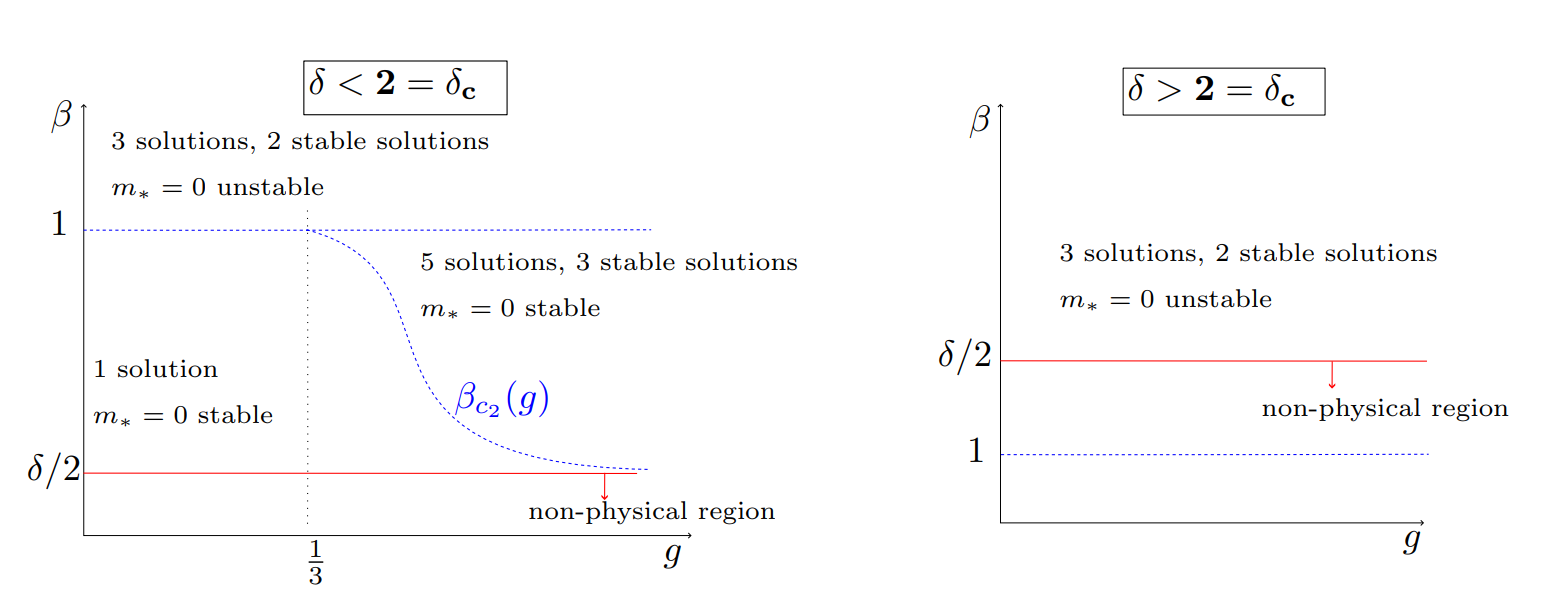}
    \caption{$(g,\beta)$ diagram for fixed $\delta$.}
    \label{g,beta diagram for fixed delta n = 0}
\end{figure}
The effect of the parameter $g$ is also visible from Figs.~\ref{g,beta diagram for fixed delta n = 0}--\ref{phase diagram beta vs delta fixed g n = 0}. For $g< g_c = 1/3$, there are either 1 or 3 solutions, and the phase diagram looks qualitatively the same as for $g=0$. There is only one solution $m_* = 0$ (paramagnetic phase) for $\beta < 1$ and three solutions for $\beta > 1$, of which the two nonzero ones are stable (ferromagnetic phase).
\\When $g> g_c = \frac{1}{3}$, the phase diagram differs with 1,3 or 5 different values for $m_*$ depending on $\beta$. Five solutions of \eqref{stationary n = 0} appear for $\beta \in [\beta_{c_2}(g), 1]$ where $\beta_{c_2}(g)$ is obtained by solving the system of equations $H(m_*) = 0, H'(m_*) = 0$ simultaneously for the pair $(\beta_{c_2}(g),m_*)$. 
The two outer ferromagnetic branches in Fig.~\ref{ms beta for different g}, as well as $m_* = 0$, are stable, while the two inner ferromagnetic branches are unstable. So by introducing $g$, a coexistence appears between ferro- and paramagnetic phases.\\ 
Furthermore, $m_* = 0$ is the only solution when $\beta < \beta_{c_2}(g)$ while for $\beta > \beta_c = 1$, three of the five solutions remain with an unstable $m_* = 0$ and stable $m_* \neq 0$. 

\subsection{\texorpdfstring{$n = 1,\ g = 0$}{n = 1, g = 0}}
For $n = 1$, temperature-dependent kinetic effects are included in the spin-flip rates \eqref{alra}, which changes the phase diagram significantly. That is clearly a nonequilibrium effect.  First, the stationary magnetization solves \eqref{general ms2}
\begin{equation}\label{ms n = 1}
    \delta \tanh{\frac{\delta}{2} \psi'(m_*)} = -2 \beta \frac{\sinh{\beta  \psi'(m_*)}-m_* \cosh{\beta  \psi'(m_*)}}{m_* \sinh{\beta  \psi'(m_*)}- \cosh{\beta  \psi'(m_*)}}, \qquad \beta \geq \frac{\delta}{2} \color{black}
\end{equation}
which explicitly depends on $\delta$. Again, for $h=0$, the magnetization $m_* = 0$ is always a solution.\\

We discuss here the case $g = 0$ of \eqref{ms n = 1}; the case $g \neq 0$ follows in the next subsection.\\
The phase diagram is given in Fig.~\ref{beta delta diagram g = 0, n = 1} with 1,3 or 5 solutions depending on the values of $\beta$ and $\delta$. That differs from the $n = 0$ case, which only has 1 or 3 solutions for $g = 0$. 
\\
Solving $H'(0) = 0$ for the stability of $m_* = 0$, shows that it is stable for $\beta \leq \beta_c(\delta)$ and unstable otherwise where
\begin{equation}\label{critical temperature n = 1}
   \beta_{c}(\delta) = \frac{1+ \sqrt{1+\delta^2}}{2} 
\end{equation}
As such, the critical temperature becomes $\delta$ dependent, increases with the degree of nonequilibrium, and reduces to the equilibrium value $\beta_c(0) = 1$. Since in equation \eqref{critical temperature n = 1}, $\beta_{c}(\delta)\geq 1$ for all $\delta$, an \textit{island of stability} opens up for the paramagnetic state, compared to the equilibrium and the $n = 0$ case. Here we thus have unstable states that become stable due to the nonequilibrium driving. 
\\ In the two limits $\delta \ll 1$ (close to equilibrium) and $\delta \gg 1$ (far from equilibrium), the critical temperatures reduce to, respectively,
\begin{equation*}
    \beta_c \approx 1 + \frac{\delta^2}{4} \text{ for } \delta \ll 1,\quad \text{ and } \qquad \beta_c \approx   \frac{1+\delta}{2} \text{ for } \delta \gg 1
\end{equation*}
{\it i.e.}, we get quadratic corrections in $\delta$ close to equilibrium while $\beta_c(\delta)$ grows linear in $\delta$ for a large nonequilibrium driving. Finally, the magnetization near the critical temperature $\beta_c(\delta)$ equals 
\begin{equation}\label{ms n = 1, g = 0}
    m_* = \pm \sqrt{\frac{24(1+\delta^{2}) (\beta-\beta_{c}(\delta))}{(4+\delta^{2})+(4 - \delta^2 ) \sqrt{1+\delta^{2}}}}  + O \left( (\beta - \beta_c(\delta))^{\frac{3}{2}} \right)
\end{equation}
The denominator in \eqref{ms n = 1, g = 0} is positive for $\delta < \delta_{c} = 2 \sqrt{2}$ and negative otherwise. 
Thus, when $\delta < 2 \sqrt{2}$, the magnetization behaves like $\sqrt{\beta -\beta_{c}(\delta)}$ for $\beta > \beta_c(\delta)$, while for $\delta > 2 \sqrt{2}$, it behaves like $\sqrt{\beta_{c}(\delta)-\beta}$ for $\beta < \beta_c(\delta)$. As before, the critical exponent for $\beta$ equals $z = \frac{1}{2}$, while $m_* \propto (\delta - \delta_c)^{-\gamma}$ with $\gamma = \frac{1}{2}$ 
\begin{figure}[H]
    \centering
    \includegraphics[scale = 0.5]{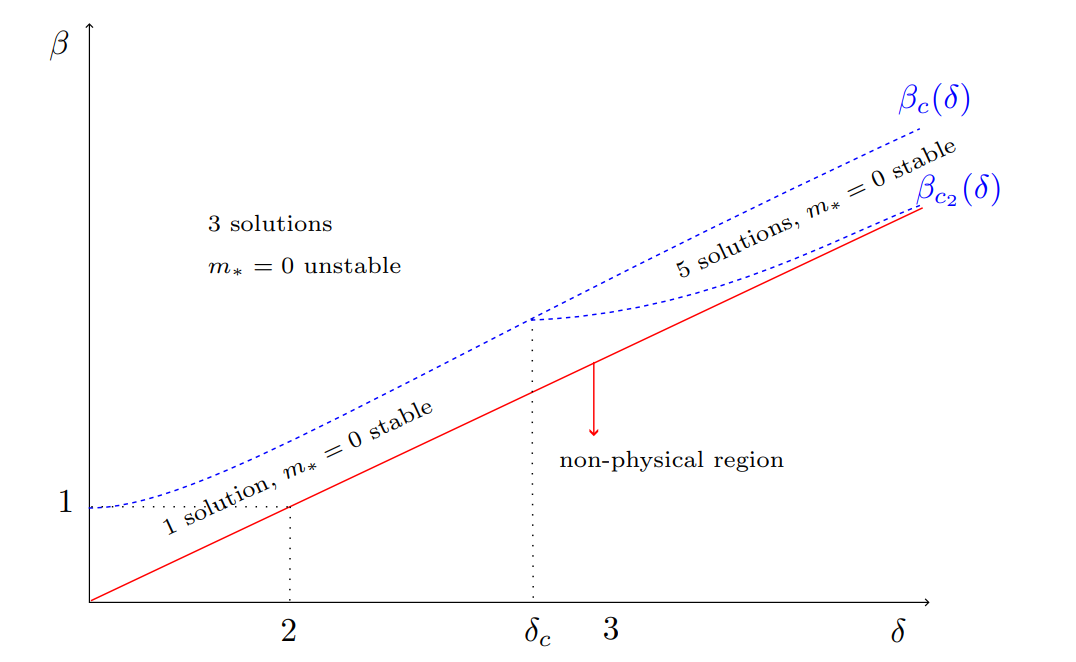}
    \caption{Phase diagram, with  $n=1$, $h =0$ and $g=0$. The strip between the upper dotted blue line and the red line is the island of stability for $m_* = 0$.}
    \label{beta delta diagram g = 0, n = 1}
\end{figure}
For $\delta < \delta_c = 2 \sqrt{2}$, the magnetization $m_* = 0$ is the only solution for $\beta < \beta_c(\delta)$ and two extra  stable ferromagnetic solutions $m_* \neq 0$ arise when $\beta > \beta_c(\delta)$. This result is similar to the $n = 0, g < \frac{1}{3}$ case from before. 
\\On the other hand, if $\delta > \delta_c = 2 \sqrt{2}$, a new critical temperature $\beta_{c_2}(\delta) \leq  \beta_{c}(\delta)$ appears for which $m_* = 0$ is the only solution when $\beta \leq \beta_{c_2}(\delta)$ (pure paramagnetic) and 5 solutions for $\beta_{c_2}(\delta) < \beta < \beta_{c}(\delta)$ (paramagentic and ferromagnetic phase coexistence).  
Two of these ferromagnetic states are stable, while $m_* = 0$ is only stable for $\beta < \beta_c(\delta)$. Note that this behaviour is very similar to the $ g > \frac{1}{3}$ case from the previous section. The major difference, however, is that the critical temperatures are $\delta$--dependent. 
\\ As before, the second critical temperature $\beta_{c_2}(\delta)$ is obtained from the equations $H(m_*) = 0, H'(m_*) = 0$. It intersects $\beta_c(\delta)$ at $\delta = \delta_c$, and moves asymptotically towards $\frac{\delta}{2}$ for large $\delta$, but it never crosses that line.

\subsection{\texorpdfstring{$n = 1, \ g \neq 0$}{n = 1, g ̸≠ 0}}
\begin{figure}[H]
    \centering
    \includegraphics[scale = 0.5]{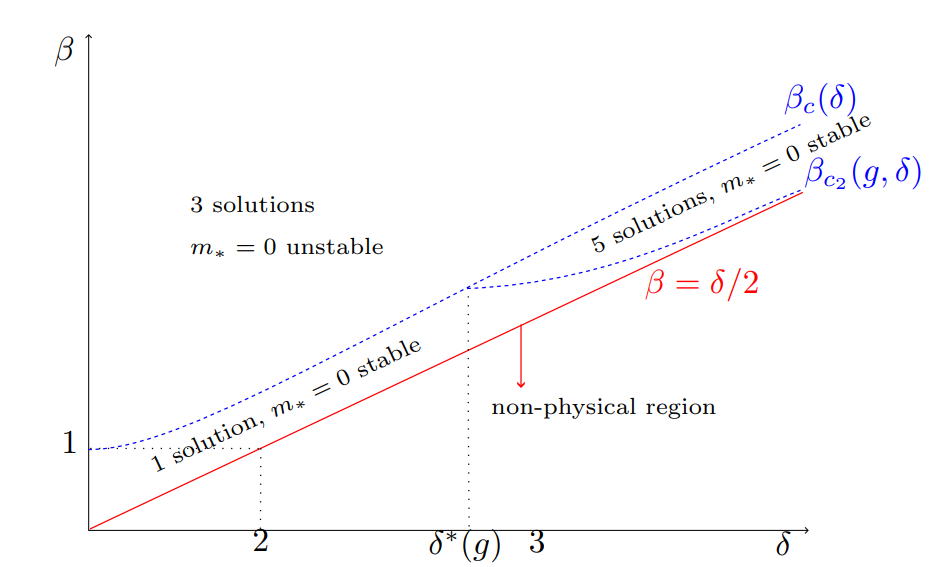}
     \caption{($\delta$, $\beta)$ diagram for fixed $g < 1/3$. The region between $\beta_{c}(\delta)$ and the red line represents the region of stability for $m_* = 0$.}
    \label{delta beta diagram g < 1/3 n = 1}
\end{figure}

\begin{figure}[H]
    \centering
    \includegraphics[scale = 0.5]{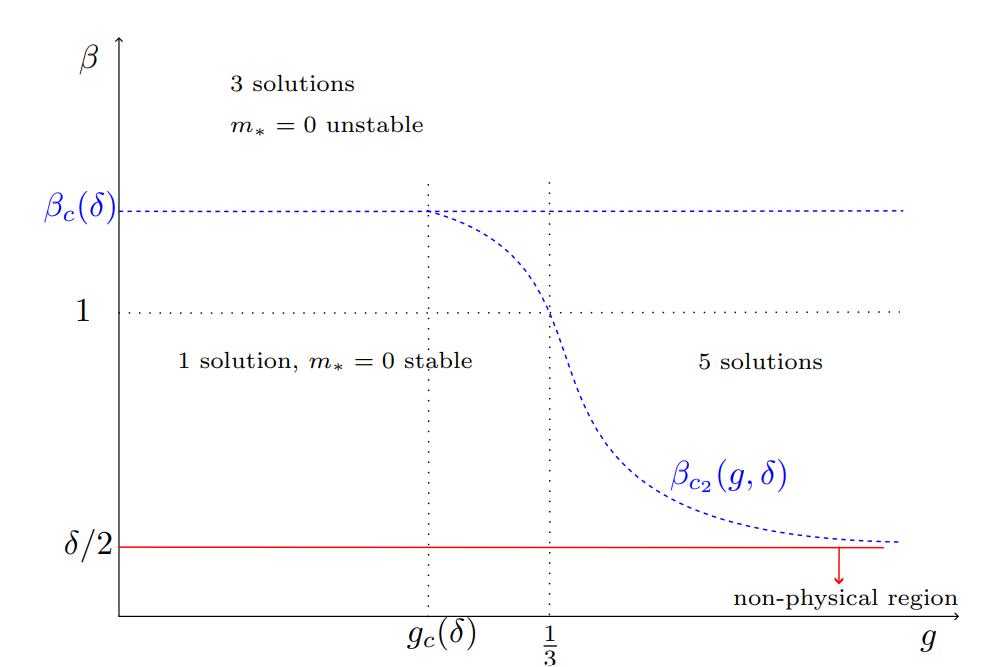}
    \caption{$(g, \beta)$ diagram for fixed $ \delta$.}
    \label{beta vs g diagram n = 1}
\end{figure}

The biggest changes for $g \neq 0$ and $n = 1$, is that the critical value $g_{c}(\delta)$ (which was the constant $\frac{1}{3}$ before) becomes $\delta$-dependent and the second critical inverse temperature $\beta_{c_2}(g,\delta)$ becomes a function of both $g$ and $\delta$. We depict the phase diagrams in Figs~\ref{delta beta diagram g < 1/3 n = 1}--\ref{beta vs g diagram n = 1}, which have a form similar to the case $n = 0, g \neq 0$.
\\As before, the critical value $g_{c}(\delta)$ follows from expanding $m_* \approx 0, \beta \approx \beta_c(\delta)$ in \eqref{ms n = 1}
\begin{equation}\label{m approx bc gc}
    m_* = \pm \sqrt{\frac{12(1+\delta^{2}) (\beta-\beta_{c}(\delta))}{6 (1+\sqrt{1+\delta^{2}})(g_{c}(\delta)-g)}} + O \left( (\beta - \beta_c(\delta))^{\frac{3}{2}} \right)
\end{equation}
with
\begin{equation}\label{g_c(delta)}
    g_{c}(\delta) = \frac{1}{3} - \frac{\delta^{2}(\sqrt{1+\delta^{2}}-1)}{12(1+\sqrt{1+\delta^{2}})}
\end{equation}
which is always smaller than $\frac{1}{3}$. The function $g_c(\delta)$ is decrease for $\delta > 0$ and becomes zero at $g_c(\delta_c) = g_c(2 \sqrt{2}) = 0$. Furthermore, for small $\delta$ (close to equilibrium), one finds
\begin{equation*}
    g_c(\delta) \approx \frac{1}{3} - \frac{\delta^4}{48} \text{ for } \delta \ll 1
\end{equation*}
{\it I.e.}, the corrections with respect to equilibrium are of fourth order in $\delta$. From \eqref{m approx bc gc},  for $g < g_{c}(\delta)$, the magnetization behaves like $\sqrt{\beta -\beta_{c}(\delta)}$ for $\beta > \beta_c(\delta)$, while for $g > g_{c}(\delta)$, it follows $\sqrt{\beta_{c}(\delta)-\beta}$ for $\beta < \beta_c(\delta)$. Furthermore, the critical exponents remain the same as before. Finally, by inverting \eqref{g_c(delta)}, the critical $\delta_c$ as a function of $g$ is obtained as
\begin{equation*}
    \delta^{*}(g) = 2 \sqrt{1-3 g+ \sqrt{1-3 g}}
\end{equation*}
The second critical temperature $\beta_{c_2}(g,\delta)$ only exists when $\delta > \delta^{*}(g)$, intersects $\beta_c(\delta)$ at $\delta^{*}(g)$ and moves asymptotically to $\frac{\delta}{2}$ as $\delta \uparrow \infty$ for all $g$.
Qualitatively, compared to the case $n=0$, the phase diagrams in Figs.~\ref{delta beta diagram g < 1/3 n = 1} and \ref{beta vs g diagram n = 1} move with $\delta$, and the critical values $\beta_{c} = 1$ and $g_{c} = 1/3$  are changed into $\delta$-dependent functions $\beta_{c}(\delta)$ and $g_{c}(\delta)$. As before, we distinguish between three different phases: pure paramagnetic $m_* = 0$ for $\beta < \beta_{c_2}(g,\delta)$, pure ferromagnetic $m_* \neq 0$ for $\beta > \beta_c(\delta)$ and a stable ferro- and paramagnetic phase in between these critical temperatures.

\section{Heat capacity}\label{section calculation heat capacity}\label{exploring the heat capacity}
This section gives the specific heat for the macroscopic two-temperature Curie-Weiss model and discusses its properties We refer to Section \ref{fhe} for the finite setup.  
The derivation of the main formula comes in Section \ref{mpe}.\\
\\
Since the heat capacity is an extensive quantity,   
$C_{ab}^{(N)} \propto N$ when $N \uparrow \infty$, the quantity of interest is the specific heat (heat capacity per spin), 
\begin{equation}\label{definition heat capacity large N}
    c_{ab} := \lim_{N \uparrow \infty} \frac{C_{ab}^{(N)}}{N} =  \lim_{N \uparrow \infty} \frac{1}{N} k_B \beta_b^2 \Big\langle \frac{\partial V^{(N)}_a}{\partial \beta_b} \Big\rangle^s_N
\end{equation}
where we have used \eqref{fheat capacity formula} for the characterization of the thermal response in terms of the quasipotential.\\
\\
{\bf Main result}:
    The specific heat  of the two-temperature Curie-Weiss model in a stable phase with magnetization $m_*$ is given by
    \begin{equation}\label{prop heat capacity}
        c_{ab} =  k_B \beta_b^2 J \frac{\partial m_*}{\partial \beta_b} \frac{ \psi''(m_*) H_a(m_*) + \psi'(m_*) H'_a(m_*)}{H'(m_*)}
    \end{equation}
We refer to \eqref{energy and psi} and \eqref{Ha} for the definitions of the functions $\psi$ and the $H_a$.  Note that in equilibrium ($\beta_1=\beta_2$), $H_a(m_*)=0$ and $2H_a= H$.
\\ Formula \eqref{prop heat capacity} is derived in Section \ref{mpe}.  We first focus on its properties and what modifications arise compared to the equilibrium situation.


\subsection{Close to equilibrium}
In \textit{equilibrium} when $J \beta_2 = J \beta_1 = J \beta_\text{eq} = \beta $, and $m_* =m_\text{eq}$,
the specific heat capacities \eqref{prop heat capacity} reduce to
\begin{equation}\label{heat capacity equilibrium first}
  c_{ab}^{\text{eq}}  =  \frac{k_B J}{2} \beta_b^2 \psi'(m_\text{eq}) \frac{\partial m_\text{eq}}{\partial \beta_b} =  \frac{k_B}{4} \beta^2 \psi'(m_\text{eq}) \frac{\partial m_\text{eq}}{\partial \beta} 
\end{equation}
in terms of the dimensionless $\beta$. There is no dependence on $a,b$, and all heat capacities agree. By adding the $c_{ab}^\text{eq}$, one gets the equilibrium result,
\begin{equation}\label{equilibrium heat capacity general}
    c_\text{eq} = 
    k_b \beta^2 \psi'(m_\text{eq}) \frac{\partial m_\text{eq}}{\partial \beta}
\end{equation}
As a reminder, for the $n = h = 0$ case, the specific heats \eqref{heat capacity equilibrium first} reduce to
\begin{equation}\label{equilibrium with g}
    c^{\text{eq}}_{ab} =\begin{cases}
			0, & \text{if $\beta < 1$}\\
            \frac{1}{8 (\frac{1}{3} - g)} k_B, & \text{if $\beta > 1$}
		 \end{cases} 
\end{equation}
where we used \eqref{near ms = 0 n = 0}. At fixed $g \neq \frac{1}{3}$, the heat capacity \eqref{equilibrium with g} makes a finite jump as $\beta \to 1$ (critical exponent $\alpha = 0$, \cite{friedli_velenik_2017}). Furthermore, still in equilibrium, a divergence of the form $(\frac{1}{3} - g)^{-\nu}$ with critical exponent $\nu = 1$  takes place at $g=1/3$. As we will see below, the critical exponents change when going to nonequilibrium. 
\\
\\ To describe the  \textit{close-to-equilibrium} behavior, we write $\beta_1 = \frac{1}{J}( \beta - \frac{\delta}{2}), \beta_2 = \frac{1}{J}( \beta + \frac{\delta}{2})$, and, for fixed $\beta$, we expand in small $\delta$
\begin{align*}
H_a & = H_\text{eq} + (-1)^{a+1} \frac{\delta}{2} \Big[ \frac{n}{\beta} H_\text{eq} - S_\text{eq} \psi'(m_*) \Big] + O(\delta^2), \qquad a=1,2 \\
J^2\beta_b^2 & = \beta^2 + (-1)^b \, \beta \delta + O(\delta^2), \qquad b = 1,2
\end{align*}
where 
\begin{eqnarray*}
    H_\text{eq}(m) &:=& \frac{2 \nu}{\beta^n} \Big(\sinh{( \beta \psi'(m))} - m\cosh{(\beta \psi'(m))} \Big),\\
    S_\text{eq}(m) &:=& \frac{2 \nu}{\beta^n} \Big(\cosh{( \beta \psi'(m))} - m\sinh{(\beta \psi'(m))} \Big)
\end{eqnarray*}
\textit{A priori}, one should also expand the stationary magnetisation $m_*$ to first order in $\delta$, but due to the symmetry \eqref{symmetry ms}, the corrections to $m_*$ are quadratic in $\delta$. Therefore, in linear order in $\delta$, there is no correction: $m_* = m_{\text{eq}}$.
That implies that to linear order in $\delta$,
\begin{align}\label{general c close to eq}
   c_{ab} = c^\text{eq}_{ab}(m_{\text{eq}}) &\Big(1 + \frac{\delta}{2} \Big[(-1)^b \frac{2}{\beta} +(-1)^{a+1} \frac{n}{\beta}+(-1)^{a} \frac{S'_\text{eq}(m_{\text{eq}}) \psi'(m_{\text{eq}})}{H'_\text{eq}(m_{\text{eq}})} \\
   & +(-1)^{a}\frac{2 S_\text{eq} \psi''(m_{\text{eq}})}{ H'_\text{eq}(m_{\text{eq}})} \Big]\Big) \nonumber 
\end{align}
In particular, 
\begin{equation}\label{only quadratic corrections}
   \sum_{a = 1}^2 \sum_{b = 1}^2 c_{ab}=  c_{11} + c_{12}+c_{21}+c_{22} = c_{11}^\text{eq} +  c_{12}^\text{eq}+ c_{21}^\text{eq} +c_{22}^\text{eq} + O(\delta^2) = c_\text{eq} + O(\delta^2)
\end{equation}
{\it i.e.}, the corrections are quadratic in $\delta$. That can be understood from \eqref{prop heat capacity} and the symmetry \eqref{symmetry ms}. Indeed, due to this symmetry, the only linear corrections in $\delta$ to $c_{ab}$  come from the $\beta_b^2$ term. Since
\begin{equation*}
    \beta_b^2 = \frac{1}{J^2} \big(\beta^2 + \beta (-1)^b \delta + O(\delta^2)\big) \qquad \sum_{a = 1}^2 \sum_{b = 1}^2 \beta_b^2 = \frac{4}{J^2} \beta^2 + O(\delta^2)
\end{equation*}
the sum $\sum_{a = 1}^2 \sum_{b = 1}^2 c_{ab}$ only has quadratic corrections in $\delta$.\\ 
In the specific case where $n =h =0$, equation \eqref{general c close to eq} reduces to
\begin{align*}
   &  c_{ab} =  c^\text{eq}_{ab}(m_{\text{eq}}) \Big(1 +\frac{\delta}{2}  \Big[(-1)^a\frac{1}{(\beta -1)}-\frac{Q_{ab}(g)}{(\frac{1}{3}-g)^2} \Big] \Big) + O \left( \beta - 1, \delta^2 \right)
\end{align*}
for some specific $Q_{ab}(g)$ where  $Q_{ab}(\frac{1}{3}) \neq 0$.

\subsection{At phase transitions}
The nonequilibrium Curie-Weiss model changes the behaviour of the specific heat near the phase transitions discussed in Section \ref{section phase diagram}.
In equilibrium, \eqref{equilibrium heat capacity general} only diverges near phase transitions due to the  $\frac{\partial m_*}{\partial \beta}$ term, whereas in nonequilibrium also the stability exponent $H'(m_*)$ in the denominator of \eqref{prop heat capacity} plays a role.\\
\\
For $n=0, g = 0, h = 0$, the phase diagram is given by Fig.~\ref{phase diagram beta vs delta fixed g n = 0}. For $\beta = J\frac{\beta_1+\beta_2}{2} \leq 1 $, the system is in the paramagnetic phase $m_* = 0$ and thus $c_{ab} = 0$, while for $\beta > 1$ in the ferromagnetic state $m_* \neq 0$
\begin{align*}
c_{11} & = c_{ab}^{\text{eq}} \cdot  (1-\frac{\delta}{2}) \left(1 +  \frac{\delta (1-\frac{\delta}{2}) }{2 (\beta -1)} + \frac{1}{32}\delta ^4-\frac{1}{16} \delta ^3+\frac{9}{8} \delta ^2-\frac{7}{4} \delta \right) +O\left(\beta -1\right) \\
    c_{12} & = c_{ab}^{\text{eq}} \cdot (1+ \frac{\delta}{2}) \left(1 +  \frac{\delta  (1+\frac{\delta}{2})}{2 (\beta
   -1)} - \frac{1}{32} \delta ^4- \frac{1}{16} \delta ^3-\frac{9}{8}
   \delta ^2 - \frac{3}{4} \delta \right) +O\left(\beta -1\right)  \\
   c_{21} & = c_{ab}^{\text{eq}} \cdot (1- \frac{\delta}{2}) \left( 1-\frac{\delta (1-\frac{\delta}{2})}{2 (\beta -1)} - \frac{1}{32}\delta ^4+\frac{1}{16} \delta ^3-\frac{9}{8} \delta ^2+\frac{3}{4} \delta \right) +O\left(\beta -1\right)\\
      c_{22} &=  c_{ab}^{\text{eq}} \cdot (1+ \frac{\delta}{2}) \left( 1-\frac{\delta  (1+ \frac{\delta}{2})}{2 (\beta
   -1)}+ \frac{1}{32} \delta ^4+ \frac{1}{16} \delta ^3+\frac{9}{8}
   \delta ^2 + \frac{7}{4} \delta \right) +O\left(\beta -1\right)  
   \end{align*}
which holds for all $\delta$ (not necessarily close to equilibrium).
The specific heats reduce to their equilibrium values when $\delta = 0$ and 
scale as $c \propto (\beta-\beta_c)^{-1}$ whenever $\delta \neq 0$. Hence, by going to nonequilibrium, a divergence of the heat capacity occurs with critical exponent $\alpha = 1$ for the $\beta$ parameter, originating from both $\frac{\partial m_*}{\partial \beta}$ and the stability exponent $H_1'(m_*) + H_2'(m_*)$:
\begin{eqnarray*}
   && \frac{\partial m_*}{\partial \beta_b} = \frac{\sqrt{3} J}{4 \sqrt{\beta - 1}} + O(1) \\
   && J\beta_b^2\frac{ \psi''(m_*) H_a(m_*) + \psi'(m_*) H'_a(m_*)}{H_1'(m_*) + H_2'(m_*)} = (-1)^{a+1} \frac{\sqrt{3} \delta (\delta + 2)^2 }{16 J \sqrt{\beta - 1}} + O(1)
\end{eqnarray*}
Note further that, for $\delta \neq 0$ and $\beta$ close enough to $1$, 
$c_{11}, c_{12}$ are positive, while $c_{21}, c_{22}$ are negative (for $\delta > 0$), which is a nonequilibrium effect. However, their sum gives
\begin{equation}\label{sum c all delta beta near 1}
    c_{11}+c_{12}+c_{21}+c_{22} 
    = c_{\text{eq}} \cdot \left(1+\frac{\delta^2}{4}\right) + O(\beta-1)
\end{equation}
which holds for all $\delta$ and is positive. Therefore, near the phase transition $\beta = 1$ for $n = 0$, the total heat capacity is larger than the equilibrium one and quadratic in $\delta$. 
\\
\\
Adding $g \neq 0$ leads to, {\it e.g.,}
\begin{align*}
    c_{12} = c^{\text{eq}}_{ab} \cdot (1 + \frac{\delta}{2}) &\Bigg( 1+\frac{\delta  (1 + \frac{\delta}{2})}{2 (\beta -1)} - \frac{1}{32 (1-3 g)}\delta ^4 
    - \frac{1}{16 (1-3 g)} \delta ^3 - \frac{3}{8} \delta ^2 \left(\frac{30 g^2-g+3}{(1-3 g)^2}\right) \\
    & - \frac{3}{4} \delta  \left(\frac{12 g^2+11 g+1}{(1-3g)^2}\right) \Bigg)
\end{align*}
with a similar form for the other $c_{ab}$. As such, the critical exponent of $\beta$ remains $\alpha = 1$, while the divergence at $g = \frac{1}{3}$ takes place with a new critical exponent $\nu = 3$ when $\delta \neq 0$ (remember that $c_{ab}^{\text{eq}} \propto (\frac{1}{3}-g)^{-1}$). 
\\
\\
Finally, for the simplest case with $n=1$, where $g = h = 0$, the heat capacities take the form
\begin{align}\label{c n = 1}
    c_{ab} = c_{ab}^{\text{eq}} \Big(1 +  \frac{\delta \ F_{ab}(\delta)}{\beta - \beta_c(\delta)} + \delta \  G_{ab}(\delta) \Big) + O\left(\beta - \beta_c(\delta) \right)
\end{align}
where $F_{ab}(\delta), G_{ab}(\delta)$ are known for general $\delta$.
The form \eqref{c n = 1} is qualitatively similar to the $n = 0$ case, but where $\beta_c$ becomes $\delta$-dependent. In particular, the critical exponent $\alpha = 1$ is unchanged.
\subsection{Low-temperature asymptotics}
We consider two possible low-temperature limits of the heat capacity evaluated in a stable magnetization. In each case, the specific heat is either zero or decays exponentially to zero, as expected from more general grounds (extended Third Law of Thermodynamics for nonequilibrium systems); see \cite{nernst_heat_example}.
\subsubsection*{Limit 1}
We consider first the low-temperature limit $\beta_1, \beta_2 \to \infty$, keeping the nonequilibrium driving $\delta = \beta_2-\beta_1$ (and hence a finite temperature difference) fixed.  Then, the specific heats go to zero exponentially fast with rate $|\psi'(\pm 1)| = \pm \psi(\pm 1)$, 
\begin{equation}\label{ca limit 1}
    c_{ab} \approx 4 k_B \frac{\beta^2 \psi'(\pm 1)^2}{(1+e^{\pm \delta \ \psi'(\pm 1)})}  e^{- 2\beta \big( \pm \psi'(\pm 1) \big)}
\end{equation}
\\
We first find the stationary magnetisation under the limit in question.
Following \eqref{general ms}, $m_*$ satisfies
\begin{equation}\label{stationary general n}
   \frac{\left( (\beta + \frac{\delta}{2})^n-(\beta - \frac{\delta}{2})^n \right)}{\left( (\beta + \frac{\delta}{2})^n+(\beta - \frac{\delta}{2})^n \right)}  \tanh{\frac{\delta}{2} \psi'(m_*)} = -\frac{\sinh{\beta  \psi'(m_*)}-m_* \cosh{\beta  \psi'(m_*)}}{m_* \sinh{\beta  \psi'(m_*)}- \cosh{\beta  \psi'(m_*)}}
\end{equation}
For all $n$, the left hand side of \eqref{stationary general n} reduces to
\begin{equation*}
    \frac{\left( (\beta + \frac{\delta}{2})^n-(\beta - \frac{\delta}{2})^n \right)}{\left( (\beta + \frac{\delta}{2})^n+(\beta - \frac{\delta}{2})^n \right)} \downarrow 0 \text{ when $\beta \uparrow \infty$ at fixed $\delta$ }
\end{equation*}
Hence, for $\beta \uparrow \infty$, the stationary magnetization solves
\begin{equation}\label{large beta magnetization equation}
    m_* = \tanh{\beta \psi'(m_*)}
\end{equation}
which is independent of $n, \delta$ and agrees with the $n = 0$ case,  \eqref{stationary n = 0}. For $h = 0$, $m_* = 0$ is always a solution to \eqref{large beta magnetization equation}, but it becomes unstable for $\beta$ large enough. Indeed, $m_* = 0$ changes stability when $H'(0) = 0$, {\it i.e.}, at $\beta = \beta_c(n,\delta)$ where
\begin{align}
     2 (\beta_{c}(n,\delta)-1) \Biggl[(\beta_{c}(n,\delta) - \delta/2)^{-n} &+ (\beta_{c}(n,\delta) + \delta/2)^{-n}\Biggr] \label{stableequation} \\
     &+ \Biggl[-(\beta_{c}(n,\delta) - \delta/2)^{-n} + (\beta_{c}(n,\delta) + \delta/
      2)^{-n}\Biggr] \delta = 0 \nonumber
\end{align}
In the limit considered, $\beta_c(n,\delta)$ is constant since $n,\delta$ are fixed quantities. As such, for $\beta > \beta_c(n,\delta)$, the zero magnetisation $m_* = 0$ is unstable. Focus thus on the nonzero solutions $m_* \neq 0$ of \eqref{large beta magnetization equation} for $\beta \uparrow \infty$. Using $\tanh{x} = \pm( 1-2 e^{\mp 2x} + o(e^{-2x}))$ for $x \uparrow \pm \infty$ leads to
\begin{equation}\label{fixed point low temperature}
    m_* = \pm( 1-2 e^{\mp 2\beta \psi'(m_*)})
\end{equation}
where the upper sign is for $m_* > 0$ and the other for $m_* < 0$. Taking $m_* = \pm(1-\varepsilon)$ with $\varepsilon \ll 1$ in \eqref{fixed point low temperature} and expanding to first order in $\varepsilon$, results in
\begin{equation*}
    \varepsilon = \frac{2 e^{\mp 2\beta \psi'(\pm 1)}}{1+ 4 e^{\mp 2\beta \psi'(1)}  \beta \psi''(\pm 1) } \approx 2 e^{\mp 2\beta \psi'(\pm 1)} 
\end{equation*}
Thus, $m_* \approx \pm( 1 - 2 e^{\mp 2\beta \psi'(\pm 1)})$ for $\beta \uparrow \infty$, and
\begin{equation}\label{large beta dm/db}
    \frac{\partial m_*}{\partial \beta} = \pm 4 \psi'(\pm 1) e^{\mp 2\beta \psi'(\pm 1)}
\end{equation}
which decreases exponentially. Furthermore,  
\begin{equation}\label{large beta c_rest}
    J k_B \beta_{j}^2\frac{ \psi''(m_*) H_a(m_*) + \psi'(m_*) H'_a(m_*)}{ H'(m_*)} \approx \pm k_B \frac{\beta^2 \psi'(\pm 1)}{J(1+e^{\pm \delta \ \psi'(\pm 1)})}
\end{equation}
Combining \eqref{large beta dm/db} and \eqref{large beta c_rest} in \eqref{prop heat capacity}, we obtain \eqref{ca limit 1}
\begin{equation*}
    c_{ab} \approx 4 k_B \frac{\beta^2 \psi'(\pm 1)^2}{(1+e^{\pm \delta \ \psi'(\pm 1)})}  e^{- 2\beta \big( \pm \psi'(\pm 1) \big)}
\end{equation*}
which decreases exponentially to $0$ with rate $\pm \psi'(\pm 1) = |\psi'(\pm 1)|$ and is independent of $n$.
\subsubsection*{Limit 2}
Another interesting limits occurs by taking $\beta_2 \to \infty$, keeping $\beta_1$ fixed, {\it i.e.}, $\beta, \delta \uparrow \infty$ keeping $\beta_1 = \frac{1}{J}(\beta-\frac{\delta}{2})$ fixed, {\it i.e.}, we only lower the temperature of the second bath. Then, the heat capacities $c_{11}, c_{12}$ and $c_{22}$ vanish, while $c_{21}$ reduces to the equilibrium result \eqref{equilibrium heat capacity general} at temperature $\beta_1$,
\begin{equation}\label{heat capacity limit 2}
c_{21} \approx k_B J \psi'(m_*) \beta_1^2 \frac{\partial m_*}{\partial \beta_1} \qquad c_{11} = c_{22} = c_{12} = 0    
\end{equation}
\\
To show, we note that analyzing \eqref{stableequation} in the limit $\delta \to \infty$ leads to
\begin{equation*}
    \beta_c(n,\delta) \propto 1 + \frac{\delta}{2}
\end{equation*}
Since $m_* = 0$ is stable for $\beta < \beta_c(n,\delta)$, it follows that for $J \beta_1 \leq 1$, $c_{ab} = 0$. In the other case $J \beta_1 > 1$, the stable magnetisation is ferromagnetic $m_* \neq 0$ and satisfies
      \begin{equation}\label{ms low temperatures}
          \frac{1\mp m_*}{1\pm m_*} = e^{\mp 2 J\beta_1 \psi'(m_*)}
      \end{equation}
    where we get the upper sign for $m_* > 0$ and the lower sign for $m_* < 0$. Equation \eqref{ms low temperatures} is obtained from \eqref{general ms} in the limit $\beta, \delta \uparrow \infty$.
    As a consequence,
      \begin{eqnarray}\label{m_* fixed beta_1}
        \frac{\partial m_*}{\partial \beta_2} &=& 0\notag\\ \frac{\partial m_*}{\partial \beta_1} &=& \frac{\pm e^{\mp 2 J \beta_1 \psi'(m_*)} J \psi'(m_*)}{\frac{1}{(1\pm m_*)^2} - e^{\mp 2 J \beta_1 \psi'(m_*)} \beta_1 J \psi''(m_*)} 
          = \frac{\pm J \psi'(m_*) (1-m_*^2)}{1 - (1-m_*^2) \beta_1 J \psi''(m_*)}
      \end{eqnarray}
Furthermore,   for \eqref{prop heat capacity} in the limit $\beta_2 \uparrow \infty$,
\begin{align}
   &\lim_{\beta_2 \to \infty} J k_B \beta_{1}^2\frac{ \psi''(m_*) H_2(m_*) + \psi'(m_*) H'_2(m_*)}{H'_1(m_*)+H'_2(m_*)} = \pm k_B J \psi'(m_*) \beta_1^2 \label{large beta c_rest 1} \\
   &\lim_{\beta_2 \to \infty} J k_B \beta_{1}^2\frac{ \psi''(m_*) H_1(m_*) + \psi'(m_*) H'_1(m_*)}{H'_1(m_*)+H'_2(m_*)} = 0 \label{large beta c_rest 2}
\end{align}
By combining \eqref{m_* fixed beta_1} and \eqref{large beta c_rest 1}-\eqref{large beta c_rest 2} in \eqref{definition heat capacity large N}, we conclude that for $J \beta_1 > 1$ and $\beta_2\uparrow\infty$,
\begin{equation}
    c_{21} \approx k_B J \psi'(m_*) \beta_1^2 \frac{\partial m_*}{\partial \beta_1} \qquad c_{11} = c_{22} = c_{12} = 0
\end{equation}
which is \eqref{heat capacity limit 2}. 
Therefore, all the excess heat by changing $\beta_1$ is produced in the second heat bath. That is physically clear as any heat will immediately flow to the zero-temperature $\beta_2 \to \infty$ bath. Furthermore, that $c_{12}, c_{22} = 0$ is a manifestation of the Third Law of Thermodynamics. \\
The form of $c_{21}$ in \eqref{heat capacity limit 2} agrees with the equilibrium result at inverse temperature $\beta_1$ as in \eqref{equilibrium heat capacity general}, but $m_*$ does not satisfy the equilibrium equation \eqref{stationary n = 0}. Furthermore, from \eqref{m_* fixed beta_1}, it follows that $c_{21}$ goes exponentially to zero with rate $|\psi'(\pm 1)|$ in case the additional limit $\beta_1 \uparrow \infty$ is taken.

\section{Computation of the specific heat matrix}\label{mpe}

\subsection{Macroscopic Poisson equation}
To derive the formula \eqref{definition heat capacity large N} for the specific heats $c_{ab}$, we need to understand the thermodynamic behaviour of the (extensive) quasipotential \eqref{fquasipotential formula}, solution of
\begin{equation}\label{pe}
L^{N}V_a\,(\sigma) = N j_{a,N}^s - P_{a,N}(\sigma) 
\end{equation}
in the limit $N \uparrow \infty$. We do that in the next paragraph by expanding both the left and right-hand side of equation \eqref{pe} to order $O \left( \frac{1}{N} \right) $. We already put here the limiting Poisson equation,
\begin{equation}\label{full equation V_a}
    \frac{S}{N} \cal V_a''(m) +\left( H(m) + \frac{\tilde{H}(m)}{N} \psi''(m) \right) \cal V_a'(m)  = - N J F_a(m) - J G_a(m) + O \left(\frac{1}{N} \right)
\end{equation}
where the smooth function ${\cal V}_a(m), m\in [-1,+1]$ is called the macroscopic quasipotential. Similar to the heat capacity, the macroscopic quasipotential is extensive, i.e.  $\cal V(m) \propto N$, which makes sure that the neglected terms are order $ O \left(\frac{1}{N} \right)$. The functions $F_a, G_a$ are given below in \eqref{fa}-\eqref{ga} while $S, \tilde{H}$ are given in \eqref{s(m) htilde}.\\
Note that the stationary dissipation $j_{a, N}^s$ in \eqref{pe} depends on a choice of stationary magnetization $m_*$. The calculation must indeed be done in a stable phase of the phase diagram, characterized by some $m_*$ which is a differentiable function of the $\beta_a$.  We fix therefore a stationary and stable magnetization $m_*$, {\it i.e.}, a solution of $H(m_*)=0, H'(m_*)<0$.\\

 \subsubsection*{Expansion of \eqref{pe}:} 
We start by writing out the right-hand side of \eqref{pe}. The heat \eqref{he1} sent to the thermal bath at inverse temperature $\beta_a$ when flipping the $k$-th spin equals
\[
q_{a,N}(\sigma,k) = \frac 1{\beta_a}\log \frac{c_a(\sigma,k)}{c_a(\sigma^k,k)} =  -2 J \left[\psi'(m^{N}(\sigma)) \sigma_k - \frac{1}{N} \psi''(m^{N}(\sigma)) + O\left(\frac{1}{N^2} \right) \right]
\]
where we use the asymptotics \eqref{ca large N}.
When the configuration is $\sigma$, the instantaneously expected heat flux (power) \eqref{hfn} to the heat bath $a$ at inverse temperature $\beta_a$ is, therefore,
\begin{align}\label{hfi}
 P_{a,N} & = \sum_k c_a(\sigma,k)\, q_{a,N}(\sigma,k) = N\,J \psi'(m^N(\sigma))\,H_a(m^N(\sigma))\\
&  + J \left[S_a(m^N(\sigma))  + \beta_a J \psi'(m^N(\sigma)) H_a(m^N(\sigma)) \right]\psi''(m^N(\sigma)) + O\left( \frac{1}{N} \right) \nonumber
\end{align}
with $H_a$ defined in \eqref{Ha} and 
\begin{align}
& S_{a}(m) := \frac{2 \nu}{(J\beta_a)^{n}}\left[ \cosh(J\beta_{a} \psi'(m))- m \,\sinh(J\beta_{a} \psi'(m))\right] \label{sa}
\end{align}
In the steady regime with stationary magnetization $m_*$, the heat current to the $a$-th reservoir, per number of spins, is \eqref{fheat current formula} and becomes
\begin{align}\label{heat current formula}
    j_{a,N}^s(m_*) &= \frac{1}{N} \langle P_{a,N} \rangle_N^s \\
    &= J \  \psi'(m_*)\,H_a(m_*)+ \frac{\psi''(m_*)}{N} \Big(S_a(m_*) + \beta_a J \psi'(m_*) H_a(m_*) \Big) + O \left(\frac{1}{N^2} \right) \nonumber
\end{align}
As a consequence, using \eqref{hfi} and \eqref{heat current formula}, when $m^N(\sigma) \to m$, the right-hand side of the macroscopic Poisson equation \eqref{pe}  becomes
\begin{align}\label{mex}
  & Nj_{a,N}^s(m_*) - P_{a,N} = - N J F_a(m) - J G_a(m) + O \left(\frac{1}{N} \right)
\end{align}
where 
\begin{align}
  F_a(m) & = \psi'(m)\,H_a(m) - \psi'(m_*^N)\,H_a(m_*^N) \\ G_a(m) & =  S_a(m) \psi''(m) - S_a(m_*^N) \psi''(m_*^N) \label{fa} \\
  & + \beta_a J \Big(\psi'(m) \psi''(m) H_a(m) - \psi'(m_*^N) \psi''(m_*^N) H_a(m_*^N) \label{ga} \Big) 
\end{align}

Next comes the left-hand side of the Poisson equation \eqref{pe}. Since all the terms on the right-hand side depend on $m^N(\sigma)$, we use $m^N(\sigma) \longrightarrow m$ and make the Ansatz $V_a = V_a(m^N(\sigma))$.  Now, for an arbitrary smooth function $f$ of $m \in [-1,1]$,
    \begin{align*}
       L^{N}f \big(m^N(\sigma) \big) = \sum_{k = 1}^{N}  c(\sigma,k) \Big[f \big(m^N(\sigma) - \frac{2}{N} \sigma_k \big)-f \big(m^N(\sigma) \big) \Big]
      \end{align*}

where $m^N(\sigma^k) = m^N(\sigma) - \frac{2}{N} \sigma_k$.
\color{black}
Using identities like
\begin{eqnarray*}
 f\big(m^N(\sigma) - \frac{2}{N} \sigma_k \big) - f\big(m^N(\sigma) \big) &=& -\frac{2}{N} f'\big(m^N(\sigma) \big) \sigma_k + \frac{2}{N^2} f''\big(m^N(\sigma) \big) + O \left(\frac 1{N^3} \right)  \\
\color{black} \hspace{-1 cm} \sum_{k = 1}^{N} e^{-J\beta_{a} \sigma_k \psi'\big(m^N(\sigma) \big)} \sigma_k&= &
-\frac{N (J\beta_{a})^n}{2 \nu} H_{a}\big(m^N(\sigma) \big)\\
\sum_{k = 1}^{N} e^{-J\beta_{a} \sigma_k \psi'\big(m^N(\sigma) \big)} &= & 
\frac{N (J\beta_{a})^n}{2 \nu} S_{a}\big(m^N(\sigma) \big)
    \end{eqnarray*}
we obtain for $m^N(\sigma) \to m$ that
\begin{equation*}
     L^{N}f\big(m^N(\sigma) \big) = H(m) f'(m) + \frac{1}{N} \Big( S(m) f''(m) +  \tilde{H}(m) f'(m) \psi''(m)\Big) + O \left(\frac{1}{N^2} \right) 
\end{equation*}
where
\begin{equation}\label{s(m) htilde}
     S(m) := S_1(m) + S_2(m) \qquad \tilde{H}(m) := J \big(\beta_1 H_1(m) + \beta_2 H_2(m) \big)
\end{equation}
\\
In the limit $N \uparrow \infty$, the solution to equation \eqref{pe} is a smooth function ${\cal V}_a(m), m\in [-1,+1]$ (macroscopic quasipotential).
Combined with the source term \eqref{mex}, we see that it verifies the macroscopic Poisson equation \eqref{full equation V_a}.

\subsection{Macroscopic quasipotential}
Defining the derivative $ W_a = \cal V_a'$, the solutions of \eqref{full equation V_a} for $W_a$ and $\cal V_a$ become
\begin{align}
    W_a(m) &= e^{\int_{\Tilde{m}}^{m} -\frac{N}{S}(H + \frac{\tilde{H}}{N} \psi'') \ \id m' }  \label{W_a}  \\
    & \Big[ W_a(\Tilde{m}) - N J \int_{\Tilde{m}}^m \id m'' \ \left(e^{\int_{\Tilde{m}}^{m''} \ \frac{N}{S}(H + \frac{\tilde{H}}{N} \psi'') \id m'} \right) \frac{1}{S} \left(F_a + \frac{G_a}{N} \right) \Big] \nonumber \\
    \cal V_a(m) &= \int_{m_*}^{m} W_a(m') \ \id m' \label{V_a}
\end{align}
where $\Tilde{m}$ is an integration constant and we used \eqref{condition V discrete}. The solution \eqref{V_a} is well-defined because the integrands are continuously differentiable everywhere and $S(m) \neq 0$ for $m \in [-1,1]$. To see this, remark that, using \eqref{sa},  $S(m) = 0$ would imply
\begin{equation}
    m = \frac{(J \beta_2)^n \cosh{J \beta_1 \psi'(m_*)} + (J \beta_1)^n \cosh{J \beta_2 \psi'(m_*)}}{
    (J \beta_2)^n \sinh{J \beta_1 \psi'(m_*)} + (J \beta_1)^n \sinh{J \beta_1 \psi'(m_*)}} > 1
\end{equation}
which is not in the integration domains $[-1,1]$. Since, furthermore, $S(0) > 0$, it holds that $S(m) > 0$ for all $m \in [-1,1] $.
\\
\\
It follows from requiring that the exponential factor be bounded for all $m$ when $N\uparrow \infty$ that $\tilde{m} \in \{0,+1,-1\}$. Indeed, the thermodynamic limit is well-defined when the integral
\begin{equation}\label{integral H}
    \int_{\Tilde{m}}^{m} \frac{1}{S} \left(H + \frac{\tilde{H}}{N} \psi''\right) \ \id m'
\end{equation}
is positive for all $m$. In the limit $N \uparrow \infty$, the sign of \eqref{integral H} is determined fully by the sign of the integral over $H$, from which it follows that, depending on the chosen temperature pair, $\tilde{m} \in \{-1,0,1 \}$. For example,  when $m_* = 0$ is unstable ($H'(0) > 0$), then the integral will be positive for all $m$ when $\Tilde{m} = 0$. Otherwise, when $m_* = 0$ is stable, the behaviour of the integrand changes, and one must take $\tilde{m} = \pm 1$.

\subsection{Specific heat calculation}
To find the heat capacity \eqref{definition heat capacity large N}, the $\beta_b$-derivative of the macroscopic quasipotential \eqref{V_a} need to be taken, which requires a careful analysis to lead us to the conclusion \eqref{prop heat capacity}.\\

We start by taking $\beta_b$ derivatives of the macroscopic quasipotential \eqref{V_a}, which gives
\color{black}
\begin{equation*}
    \frac{\partial \cal V_a}{\partial \beta_b} = W_a(m_*) \  \frac{\partial m_*}{\partial \beta_b} + \int_{m_*}^{m} \frac{\partial W_a}{\partial \beta_b} \ \id m' 
\end{equation*}
with
\begin{align*}
    & \frac{\partial W_a}{\partial \beta_b} = e^{\int_{\tilde{m}}^{m} -\frac{N}{S}(H + \frac{\tilde{H}}{N} \psi'') \ \id m' } \Big\{ \frac{\partial W_a(\tilde{m})}{\partial \beta_b}   - W_a(\tilde{m}) \int_{\tilde{m}}^m \id m' \ \frac{\partial}{\partial \beta_b} \Big[ \frac{N}{S}(H + \frac{\tilde{H}}{N} \psi'') \Big]  \\
    &+ NJ \int_{\tilde{m}}^m \id m' \ \frac{\partial}{\partial \beta_b} \Big[ \frac{N}{S}(H + \frac{\tilde{H}}{N} \psi'') \Big]   \cdot \int_{\tilde{m}}^m \id m'' \ \Big[ e^{\int_{\tilde{m}}^{m''} \frac{N}{S}(H+ \frac{\tilde{H}}{N} \psi'') \ \id m'} \frac{1}{S} \left(F_a + \frac{G_a}{N} \right) \Big] \\
    & - N J \int_{\tilde{m}}^m \id m'' \ \frac{\partial}{\partial \beta_b} \Big[e^{\int_{\tilde{m}}^{m''} \frac{N}{S}(H + \frac{\tilde{H}}{N} \psi'') \ \id m'} \frac{1}{S} \left(F_a + \frac{G_a}{N} \right)\Big] \Big\}
\end{align*}
This derivative is analytic everywhere on its integration domain such that in the limit $m \to m_*$, the integral $\int_{m_*}^{m} \frac{\partial W_a}{\partial \beta_b} \ \id m'$ does not contribute. Therefore,
\begin{equation}\label{derv}
\frac{\partial \cal V_a}{\partial \beta_b}(m_*) = W_a(m_*) \frac{\partial m_*}{\partial \beta_b}
\end{equation}
Note further that in the limit $N \uparrow \infty$, \eqref{W_a} and \eqref{V_a} reduce to
\begin{align*}
   & w_a(m) = \lim_{N \uparrow \infty} \frac{W_a(m)}{N} 
   = - J \frac{\psi'(m) H_a(m) - \psi'(m_*) H_a(m_*)}{H(m)} \ \text{for $m \neq \Tilde{m}$} \\
   & v_a(m) = \lim_{N \uparrow \infty} \frac{\cal V_a(m)}{N} = - J  \int_{m_*}^m  \frac{\psi'(m) H_a(m) - \psi'(m_*) H_a(m_*)}{H(m)} 
\end{align*}
and
\begin{align}\label{limit was(s)}
  &  w_a(m_*) 
  = \lim_{m \to m_*} - J \frac{\psi'(m) H_a(m) - \psi'(m_*) H_a(m_*)}{H(m)} \nonumber \\
  & = - J \frac{\psi''(m_*) H_a(m_*) + \psi'(m_*) H_a'(m_*)}{H'(m_*)} \ \text{for $m_* \neq \Tilde{m}$} 
\end{align}
Taking $N \uparrow \infty$ and using \eqref{limit was(s)}, we conclude that 
\begin{equation}\label{spch}
\lim_N \frac{1}{N} \frac{\partial \cal V_a}{\partial \beta_b}= J \frac{\partial m_*}{\partial \beta_b} \frac{ \psi''(m_*) H_a(m_*) + \psi'(m_*) H'_a(m_*)}{H'(m_*)}
\end{equation}
Note that applying \eqref{limit was(s)} only holds for $m_* \neq \tilde{m} \in \{-1,0,1 \}$, but since $W_a(m_*)$ is multiplied by $\frac{\partial m_*}{\partial \beta_b}$ in \eqref{derv}, which is zero for $m_* \in \{-1,0,1 \}$, the last expression is valid for all $m_*$. From \eqref{spch}, the main formula for the specific heat \eqref{prop heat capacity} follows
\begin{align}
    c_{ab}(m_*) &=  \lim_N \frac{1}{N} k_B \beta_b^2 \Big\langle \frac{\partial \cal V_a}{\partial \beta_b}(m_*, \sigma) \Big\rangle^s_N \nonumber \\
  & = k_B \beta_b^2 J \frac{\partial m_*}{\partial \beta_b} \frac{ \psi''(m_*) H_a(m_*) + \psi'(m_*) H'_a(m_*)}{H'(m_*)}
\end{align}

 We finally observe that, by the symmetry of the reservoirs,
\begin{equation*}
    c_{11} c_{22} = c_{12} c_{21}
\end{equation*}
implying that the determinant of the specific heat matrix vanishes. One can indeed not always reconstruct the temperature changes from observing the excess heat.

\section{Summary and outlook}
The Curie-Weiss model is turned into a nonequilibrium spin model by randomly switching between thermal baths. The phase diagrams change with the degree of nonequilibrium, also depending significantly on the temperature dependence of the time-symmetric part in the spin-flip rates. Most notably, new regions of stability may arise for otherwise unstable phases. Furthermore, the critical temperature moves with the nonequilibrium amplitude.\\
The nonequilibrium specific heat is obtained in the thermodynamic limit $N \uparrow \infty$ following the excess heat formalism. In contrast with the standard 
Curie-Weiss model, where the specific heat shows a jump at the critical temperature, under the two-temperature driving, it diverges at the (also new) phase transitions 
with critical exponent $\alpha = 1$. The low-temperature asymptotics satisfies an extended Nernst postulate, and the specific heat vanishes exponentially fast at absolute zero.\\
\\
We believe that the two-temperature Curie-Weiss model is the first to combine a systematic study of the nonequilibrium phase diagram with an analysis of thermal response in 
terms of heat capacity. 
Other response functions can certainly be studied as well. On the other hand, our model is slightly artificial and does not directly lead to a more general Landau-type view on 
the influence of nonequilibrium aspects on phase transitions \cite{criticalexponents}. As a matter of fact, and in contrast with equilibrium, different versions of 
mean-field modelling may exist, possibly giving quite different results.\\
Extensions of our work indeed include a Landau field theory approach, taking a finite switching rate $r$ between reservoirs as in \eqref{finite r} and including kinetically-different 
thermal baths. They make the subjects of future work.\\

\noindent {\bf  Acknowledgment:}  The two-temperature Curie-Weiss model of the present paper was first proposed by Karel Netočný to study the nonequilibrium phase diagram. We are grateful for his suggestion.
\\ Furthermore, the authors thank the reviewers for their thoughtful comments and useful suggestions.
\\
\\
{\bf Data Availability:} Data sharing is not applicable to this article as no data sets were generated or analyzed during
the current study.
\\
\\
{\bf Conflict of interest} The authors have no relevant financial or non-financial interests to disclose.

\newpage
\appendix
\section{Example: \texorpdfstring{$N = 2$}{N = 2}}\label{two}
\begin{figure}[H]
    \centering
    \includegraphics[scale = 0.65]{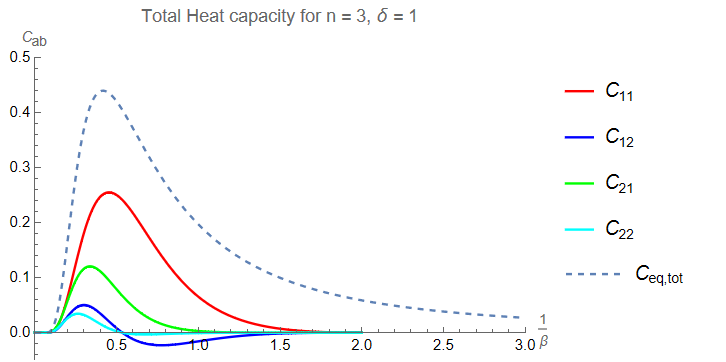}
    \caption{Heat capacities vs $\frac{1}{\beta} = \big( J_g \frac{\beta_1 + \beta_2}{2} \big)^{-1}$ for $n = 3$ and $\delta = J_g \big( \beta_2 - \beta_1 \big) = 1$. The graphs stop at $\beta = \frac{\delta}{2}$ where $\beta_1 = 0$. We plot the total equilibrium result $C_{\text{eq, tot}}$ with a dashed line. (Made using Mathematica version 13.1.0.0 \cite{Mathematica}.)}
    \label{heat capacity n = 3}
\end{figure}

When the system consists of two spins, the magnetization becomes $m^{N = 2} = \frac{\sigma_1+\sigma_2}{2} \in \{-1,0,1\}$, 
and energy $E(\sigma) =  -J_g\frac{1+\sigma_1 \sigma_2}{2} \in \{0,-J_g\}$, where $J_g = J \big( 1 + \frac{g}{2} \big)$. Therefore, the process reduces to a  two-level switch where $g$ only affects the energy difference. \\
\\
First, the heat fluxes \eqref{hfn}, $P_{a,N=2}(\sigma)$ become
\begin{align*}
    P_{a,N}(1,1) =  P_{a,N}(-1,-1) & = -\frac{2J_g \nu}{(J_g\beta_{a})^n} e^{- \beta_{a} \frac{J_g}{2}},  \\
     P_{a,N}(1,-1) = P_{a,N}(-1,1) &= \frac{2J_g \nu}{(J_g\beta_{a})^n} e^{\beta_{a} \frac{J_g}{2}}
\end{align*}
The quasipotential $V_a^{(N=2)}$ in \eqref{fquasipotential formula} takes the form $V_a^{(N = 2)}= (V_{a1}, V_{a2}, V_{a2}, V_{a1})$ with
\begin{align*}
  V_{a1} & = - v_{ab} \ e^{-\frac{1}{2} \left(2 \beta_a+\beta_b\right) J_g}  \left(e^{\frac{\beta_a J_g}{2}} \beta_a^n+e^{\frac{\beta_b J_g}{2}} \beta_b^n\right)
    \\
    V_{a 2} &= v_{ab} \ e^{-\frac{\beta_a J_g}{2}} \left(e^{\frac{\beta_b J_g}{2}} \beta_a^n+e^{\frac{\beta_a J_g}{2}} \beta_b^n\right) \\
     v_{ab} & = J_g \left(e^{\beta_a J_g}+1\right) \beta_b^n \left(2 \beta_1^n \cosh \left(\frac{\beta_2 J_g}{2}\right)+2 \beta_2^n \cosh \left(\frac{\beta_1 J_g}{2}\right)\right)^{-2}
\end{align*}
where $a \neq b$. That, from \eqref{fheat capacity formula}, leads to the heat capacities,
\begin{align*}
& C_{aa}^{(N = 2)} =k_B \ \tilde{C}_{ab} \ \beta_a \ \beta_b^{2 n} \left(J_g \beta_a \beta_b^n + J_g \cosh{\frac{J_g}{2} (\beta_a-\beta_b)} - 2 \ n \ \beta_a^n \sinh{\frac{J_g}{2} (\beta_a-\beta_b)} \right) \\
  &C_{ab}^{(N = 2)} = k_B \ \tilde{C}_{ab} \ \beta_a^n \ \beta_b^{n+1} \left(J_g \beta_a^n \beta_b + J_g \beta_b^{n+1} \cosh{\frac{J_g}{2} (\beta_a-\beta_b)} + 2 \ n \ \beta_b^n \sinh{\frac{J_g}{2} (\beta_a-\beta_b)} \right)\\
  &\tilde{C}_{a b} = \frac{J_g}{4} \cosh{\frac{J \beta_a}{2}} \left(\beta_1^n \cosh \left(\frac{\beta_2 J_g}{2}\right)+ \beta_2^n \cosh \left(\frac{\beta_1 J_g}{2}\right)\right)^{-3} 
\end{align*}
where again $a \neq b$. In equilibrium $\beta_2 = \beta_1$, all of them reduce to
\begin{equation*}
    C_{\text{eq,}ab}^{(N = 2)} = k_B\frac{\beta_1^2 J_g^2 e^{\beta_1 J_g}}{4 \left(e^{\beta_1 J_g}+1\right)^2} 
\end{equation*}
which is a quarter of the total equilibrium heat capacity $C_\text{eq,tot}$ for a two-level system  \cite{gopal2012specific}.\\
The heat capacities are plotted in Fig. \ref{heat capacity n = 3}, where some are seen to obtain negative values, here for $n = 3$. As discussed in \cite{dolai2023manybody}, this happens due to a negative correlation between the
quasipotential $V_a^{(N)}$ and the change in stationary distribution $\rho^s$.

\bibliography{CW}

\end{document}